%% file: main.tex
\documentclass[sigconf, screen]{acmart}

\usepackage[table]{xcolor}
\definecolor{cvprblue}{rgb}{0.21,0.49,0.74}

\hypersetup{
  colorlinks=true,
  linkcolor=cvprblue,
  citecolor=cvprblue,
  urlcolor=cvprblue
}

\AtBeginDocument{%
  }

\setcopyright{acmlicensed}

\copyrightyear{2026}
\acmYear{2026}
\setcopyright{cc}
\setcctype{by}
\acmConference[KDD '26]{Proceedings of the 32nd ACM SIGKDD Conference on Knowledge Discovery and Data Mining V.2}{August 09--13, 2026}{Jeju Island, Republic of Korea}
\acmBooktitle{Proceedings of the 32nd ACM SIGKDD Conference on Knowledge Discovery and Data Mining V.2 (KDD '26), August 09--13, 2026, Jeju Island, Republic of Korea}
\acmDOI{10.1145/3770855.3818092}
\acmISBN{979-8-4007-2259-2/2026/08}

\usepackage{subfigure}
\newcommand{\cm}{\mathcal}
\newtheorem{theorem}{Theorem}
\newtheorem{thm}{Theorem}
\newtheorem{assumption}[thm]{Assumption}
\newtheorem{definition}[theorem]{Definition}
\newtheorem{lemma}[theorem]{Lemma}

\usepackage{multirow}

\usepackage[edges]{forest} 
\usetikzlibrary{graphs}
\usetikzlibrary{arrows.meta}

\usepackage{algorithm}
\let\classAND\AND
\let\AND\relax
\usepackage{algorithmic}

\let\AND\classAND
\AtBeginEnvironment{algorithmic}{\let\AND\algoAND}

\begin{document}


\title{Local Clustering on Complex Graphs and Complex Hypergraphs}

\author{Zihao Li}
\affiliation{%
  \institution{University of Illinois Urbana-Champaign}
  \state{Illinois}
  \country{USA}
}
\email{zihaoli5@illinois.edu}

\author{Dongqi Fu}
\affiliation{%
  \institution{Meta}
  \state{California}
  \country{USA}
}
\email{dongqifu@meta.com}

\author{Hengyu Liu}
\affiliation{%
  \institution{University of Illinois Urbana-Champaign}
  \state{Illinois}
  \country{USA}
}
\email{hengyu2@illinois.edu}

\author{Jingrui He}
\affiliation{%
  \institution{University of Illinois Urbana-Champaign}
  \state{Illinois}
  \country{USA}
}
\email{jingrui@illinois.edu}


\renewcommand{\shortauthors}{Li et al.}


\addtocontents{toc}
{\protect\setcounter{tocdepth}{-1}}

\begin{abstract}
  Local/seeded clustering aims to find a compact cluster near the given starting instances. While most existing studies on graph clustering assume a discrete graph setting (i.e., unweighted, undirected graphs without self-loops), real-world graphs can be more complex. In this paper, we extend the classic non-approximating Andersen-Chung-Lang (ACL) clustering algorithm beyond discrete graphs and generalize its quadratic optimality to a wider range of complex graphs, including \underline{weighted, directed, and self-looped graphs} and \underline{hypergraphs with edge-dependent vertex weights}. Specifically, by leveraging PageRank, we propose two algorithms: \textbf{GeneralACL} for graphs and \textbf{HyperACL} for hypergraphs. We prove that, under two mild conditions, both algorithms can identify a quadratically optimal cluster in terms of conductance\footnote{The conductance of the returned cluster $\Phi$ and the optimal conductance $\Phi^*$ satisfy $\Phi \leq O(\sqrt{\Phi^*})$.} with at least $\frac{1}{2}$ probability. Additionally, we provide experiments to validate our theoretical findings. Our code is available at \url{https://github.com/iDEA-iSAIL-Lab-UIUC/HyperACL}.
\end{abstract}




\maketitle

\section{Introduction}

\begin{figure}[t]
    \centering
    \includegraphics[width=0.50\textwidth]{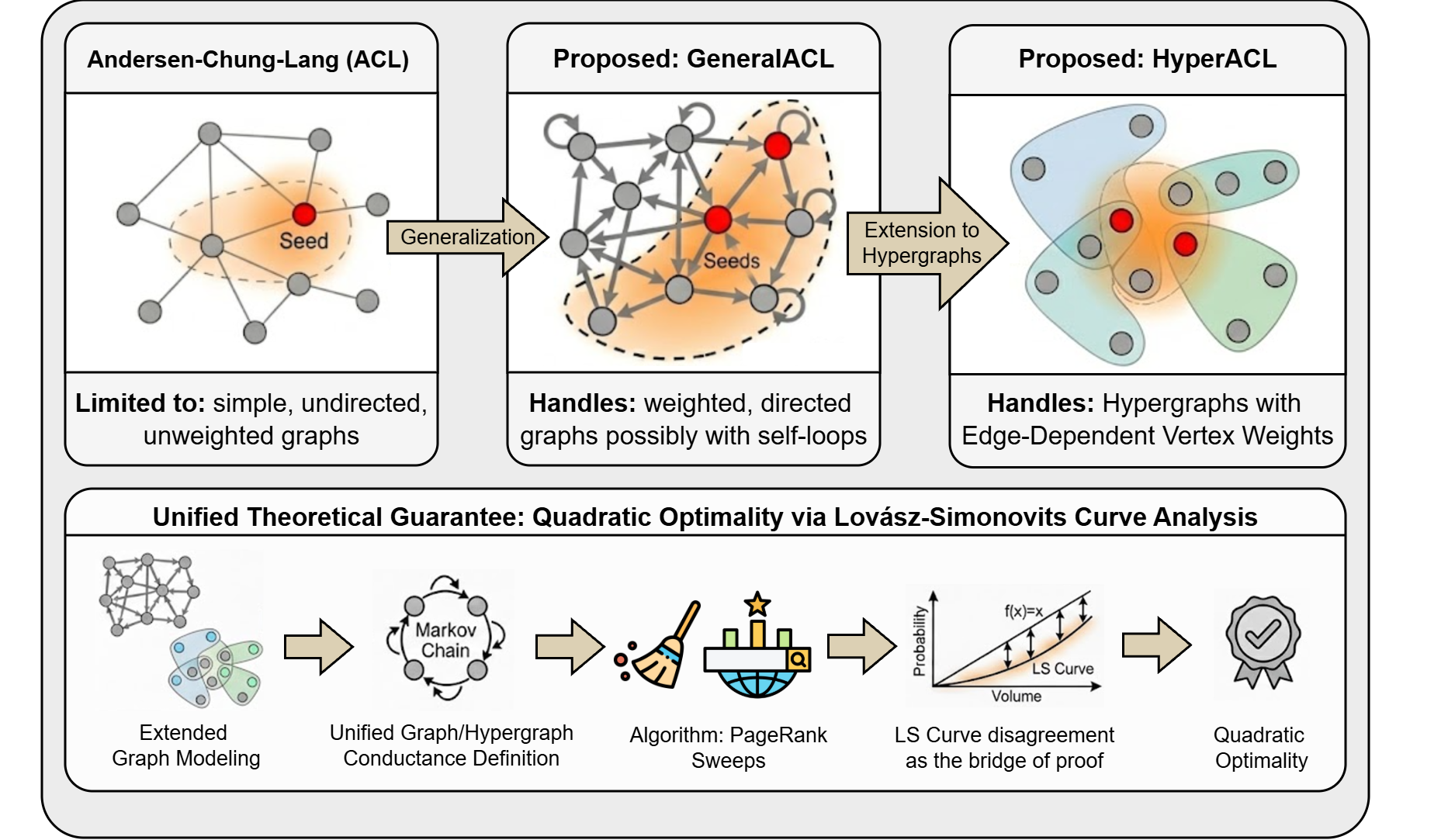}
    \caption{Comparison among ACL and our proposed GeneralACL and HyperACL, and the corresponding theoretical guarantee of quadratic optimality for both methods.}
    \label{fig: teasing}
\end{figure}

Local/seeded graph clustering is a fundamental research problem to identify a single cluster around a given seed node or set of seed nodes \citep{DBLP:journals/corr/abs-1810-07324, DBLP:conf/cvpr/TolliverM06, DBLP:conf/cikm/AlvarezYKI13, DBLP:conf/rskt/KardanR18}.
However, most existing research works assume the graphs to be undirected and unweighted \citep{DBLP:conf/focs/AndersenCL06, DBLP:conf/icml/VeldtGM16, DBLP:journals/mp/FountoulakisRSC19, DBLP:conf/icdcs/ZhangJHGYY21, DBLP:journals/siamcomp/SpielmanT13}. While a few research works extend to directed graphs \citep{DBLP:journals/im/AndersenCL08}, real-world graphs could be more complex, as the vertices and edges in the graphs can have individual weights \citep{DBLP:conf/soda/HenzingerLRW24, DBLP:conf/kdd/GuZCP21, DBLP:conf/focs/DuanMSY23}, and the graphs may contain self-loops \citep{DBLP:journals/ploscb/SharmaYT23, DBLP:conf/wivace/MereloM23}. 
Converting such graphs to undirected forms discards important information, while directly designing clustering algorithms for them remains an open problem.
Previously, a classic work adopts the PageRank \citep{page1999pagerank} to produce a quadratically optimal local clustering algorithm (i.e., \textbf{A}ndersen-\textbf{C}hung-\textbf{L}ang algorithm) \citep{DBLP:conf/focs/AndersenCL06} on undirected and unweighted graphs without self-loops. Specifically, this ACL algorithm~\citep{DBLP:conf/focs/AndersenCL06} sweeps over the personalized PageRank vector and returns the local cluster with the smallest conductance.
Later on, a seminal work~\citep{DBLP:journals/im/AndersenCL08} extends the ACL algorithm and allows the input graphs to be directed. 

In this work, we first propose \textbf{GeneralACL}, which further generalize the ACL algorithm to any graph that is possibly weighted, directed, and even with self-loops. Moreover, GeneralACL allows more than one starting vertex (seed node) in the typical ACL, which breakthrough fully releases the effectiveness and potential of PageRank, by using a single multi-seed computation in one-shot to save the compute of multiple single-seed runs. The only assumption of our GeneralACL is easy to acquire, i.e., a well-defined graph random walk with the existence of the stationary distribution, as illustrated in Section~\ref{section: local}.
Additionally, during the sweeping process of GeneralACL, we introduce an early-stop mechanism to make the algorithm,  apart from the PageRank computation, strongly local, i.e., the runtime can be controlled by the size of the output cluster rather than the size of the graph. 

\renewcommand\arraystretch{1.3}
\begin{table*}[t]
\caption{Table of notation and hyperparameters for quick reference}
\label{tab: notation}
\centering
\scalebox{0.85}{
\begin{tabular}{|p{80pt}|p{450pt}|}
\hline Symbol&Definition and Description\\
\hline
\hline$\cm{G} = (\cm{V}, W, \varphi)$ & graph being investigated, with vertex set $\mathcal{V}$, adjacency matrix $W$, vertex weight mapping $\varphi$\\
\hline$\mathcal{H} = (\mathcal{V}, \mathcal{E}, \omega, \gamma)$ & hypergraph being investigated, with vertex set $\mathcal{V}$, hyperedge set $\mathcal{E}$, edge weight mapping $\omega$ and edge-dependent vertex weight mapping $\gamma$\\
\hline$n = |\cm{V}|$ & number of vertices\\
\hline$m$ & number of hyperedge-vertex connections  in Hypergraph $\cm{H}$, $m = \sum_{e\in \cm{E}} |e|$\\
\hline$d(v)$ & degree of vertex $v$, $d(v) = \sum_{e\in E(v)}w(e)$. $w(e)$ is hyperedge weight mappings (Definition \ref{df: edvw hypergraph})\\
\hline$\delta(e)$ & degree of hyperedge $e$, $\delta(e) = \sum_{v \in e}\gamma_e(v)$. $\gamma_e(v)$ is vertex weight mappings (Definition \ref{df: edvw hypergraph})\\
\hline$R$ & $|\cm{E}| \times |\cm{V}|$ vertex-weight matrix (Definition \ref{df: R, W, Dv, De})\\
\hline$W$ & $|\cm{V}| \times |\cm{E}|$ hyperedge-weight matrix (Definition \ref{df: R, W, Dv, De})\\
\hline$D_\cm{V}$ & $|\cm{V}| \times |\cm{V}|$ vertex-degree matrix (Definition \ref{df: R, W, Dv, De})\\
\hline$D_\cm{E}$ & $|\cm{E}| \times |\cm{E}|$ hyperedge-degree matrix (Definition \ref{df: R, W, Dv, De})\\
\hline$P$ & $|\cm{V}| \times |\cm{V}|$ transition matrix of random walk on $\cm{H}$ (Definition \ref{df: transition matrix})\\
\hline$\phi$ & $1 \times |\cm{V}|$ stationary distribution of random walk (Definition \ref{df: stationary distribution})\\
\hline$\Pi$ & $|\cm{V}| \times |\cm{V}|$ diagonal stationary distribution matrix (Definition \ref{df: stationary distribution})\\
\hline$p$ & $1 \times |\cm{V}|$ probability distribution on $\cm{V}$ (Definition \ref{df: p(S)})\\
\hline$\alpha$ & restart probability of random walks (Definition \ref{df: pagerank})\\
\hline$\partial$ & boundary notation (Definition \ref{df: volume of boundary})\\
\hline$vol$ & volume notation (Definition \ref{df: volume of a set})\\
\hline$pr(\alpha, s)$ & lazy personalized PageRank vector with restart probability $\alpha$ and stochastic vector $s$ (Definition \ref{df: pagerank})\\
\hline
\end{tabular}}
\label{TB: Notations}
\end{table*}
\renewcommand\arraystretch{2}

One direct application of our GeneralACL algorithm is to solve local clustering problems over \textit{hypergraphs}~\citep{DBLP:journals/csur/AntelmiCPSSY24}, where each edge is able to connect more than two vertices. Through modeling group interactions, hypergraphs benefit a wide array of applications \citep{DBLP:journals/corr/abs-1809-09401, DBLP:conf/nips/YadatiNYNLT19, DBLP:journals/jim/XiaZHL22, DBLP:journals/kbs/WuWSFZY21, grilli2017higher, DBLP:journals/tois/HuangCYHZ20, DBLP:conf/smartcom/LiuHZLXH22}.
However, to the best of our knowledge, there is no hypergraph local clustering algorithm that is proven to have any quadratic optimality.
Based on our GeneralACL, we also propose \textbf{HyperACL}, which further extends the general hypergraph modeling with the recent proposed \textbf{e}dge-\textbf{d}ependent \textbf{v}ertex \textbf{w}eights (EDVW) hypergraph~\citep{li2024hypergraphs}, where nodes from a single hyperedge can have different weight, formal definition is given in Definition \ref{df: edvw hypergraph}. Inheriting from GeneralACL, HyperACL can, with at least $\frac{1}{2}$ probability, find a quadratically optimal local cluster in terms of our proposed hypergraph conductance, as illustrated in Definition \ref{df: conductance}. Empirically, extensive experiments on diverse datasets confirm both the efficiency and effectiveness of our algorithms.

\section{Preliminaries}
\label{sec:preliminaries}

We use calligraphic letters (e.g., $\mathcal{A}$) for sets, capital letters for matrices (e.g., $A$), and unparenthesized superscripts to denote the power (e.g., $A^{k}$).
For matrix indices, we use $A_{i, j}$ or $A(i, j)$ interchangeably to denote the entry in the $i^{th}$ row and the $j^{th}$ column. For row vector or column vector $v$, we use $v(i)$ to index its $i^{th}$ entry.
We denote graph as $\mathcal{G}$ and hypergraph as $\mathcal{H}$. For quick reference, we provide a table of notation (Table \ref{tab: notation}) and logical flow (Figure \ref{fig: technical map}).

\subsection{Graphs, Possibly Edge-weighted, Node-weighted, Directed and Self-looped}
\begin{definition}\label{df: graph} (Graph).
A graph $\cm{G} = (\cm{V}, W, \varphi)$ consists of a set of vertices $\cm{V}$, a (generalized) adjacency matrix $W \in \mathbb{R}_{\geq 0} ^ {|\cm{V}| \times |\cm{V}|}$, vertex weights $\varphi(v): \cm{V} \rightarrow \mathbb{R}_{+}$ on every vertex $v \in \cm{V}$. $W_{i, j} = 0$ indicates no edge from \(i\) to \(j\). Without loss of generality, we index the vertices by $1, 2, ..., |\cm{V}|$, and let $\cm{V} = \{1, 2, ..., |\cm{V}|\}$.
\end{definition}



A graph is \textbf{directed} if and only if there exists \( i, j \in \cm{V} \) such that $W_{i, j} \neq W_{j, i}$. 
A graph is \textbf{self-looped} if and only if there exists at least one vertex \( v \in \cm{V} \) such that $W_{v, v} > 0$. 
We allow the graphs to have arbitrary edge weights and node weights to model complex systems. Yet we assume the graph is strongly connected for the existence of stationary distribution for random walk.


\begin{definition}
\label{df: graph random walk}
 \citep{tong2006fast} (Graph random walk).
A \textit{random walk} on a graph is a Markov Chain on $\cm{V}$ with transition probabilities matrix $P$. The \textit{stationary distribution} of the random walk with transition matrix $P$ is a $ 1 \times|\cm{V}|$ row vector $\phi$ such that
\begin{equation}
    \phi P = \phi;~ \phi(u) > 0 \,\, \forall u\in \cm{V};~ \sum_{u\in{\cm{V}}}\phi(u) = 1
\end{equation}
\end{definition}
For discrete graph setting, the transition matrix $P=D^{-1}W$ where $D$ is the diagonal degree matrix $D_{i, i} = \sum_{k=1}^{|\mathcal{V}|} W_{i, k}$. 
We assume there is a well-defined irreducible and aperiodic transition matrix \( P \) that satisfies the properties of a row-stochastic matrix. In fact, an irreducible \(P\) is granted by being strongly connected.

\begin{assumption}
\label{assumption: well-defined transition matrix}
    We assume the existence of an irreducible and aperiodic transition matrix \( P \in \mathbb{R}_{\geq 0}\) that is row-stochastic (row sum 1). This makes sure a unique stationary distribution \( \phi \) satisfying \( \phi P = \phi \).
\end{assumption}

\input{logical_flow}

\subsection{Hypergraphs, with Possibly Edge-dependent Vertex Weights}
\subsubsection{EDVW Hypergraph Formulation}
A hypergraph consists of vertices and hyperedges. A hyperedge $e$ is a connection between two or more vertices. We use the notation $v \in e$ if the hyperedge $e$ connects vertex $v$, a.k.a., ``$e$ is incident to $v$".
Edge-dependent vertex weights (EDVW) modeling \citep{DBLP:conf/icml/ChitraR19} is one of the most generalized modeling methods of hypergraphs, defined as follows. 
\begin{definition}\label{df: edvw hypergraph} \citep{DBLP:conf/icml/ChitraR19} (EDVW hypergraph definition).
A hypergraph $\cm{H} = (\cm{V}, \cm{E}, \omega, \gamma)$ with edge-dependent vertex weight is defined as a set of vertices $\cm{V}$, a set $\cm{E} \subseteq 2^\cm{V}$ of hyperedges, a weight mapping $\omega(e): \cm{E} \rightarrow \mathbb{R}_+$ on every hyperedge $e \in \cm{E}$, and weight mappings $\gamma_e(v): \cm{V} \rightarrow \mathbb{R}_{\geq 0}$ corresponding to $e$ on every vertex $v$. For $e_1 \neq e_2$, $\gamma_{e_1}(v)$ and $\gamma_{e_2}(v)$ may be different. Without loss of generality, we index the vertices by $1, 2, ..., |\cm{V}|$, and let $\cm{V} = \{1, 2, ..., |\cm{V}|\}$.
\end{definition}


For instance, in a citation hypergraph, each publication is captured by a hyperedge. While each publication may have different citations (i.e., edge-weight $w(e)$), each author may have an individual weight of contributions (i.e., publication-dependent $\gamma_e(v)$). 


\begin{definition}
\label{df: R, W, Dv, De}
\citep{DBLP:conf/icml/ChitraR19} (matrices of an EDVW hypergraph).
$E(v) = \{e \in \cm{E} \, s.t. \, v \in e\}$ is the set of hyperedges incident to vertex $v$. $d(v) = \sum_{e\in E(v)}w(e)$ denotes the degree of vertex $v$. $\delta(e) = \sum_{v \in e}\gamma_e(v)$ denotes the degree of hyperedge $e$. The \textit{vertex-weight matrix} $R$ is an $|\cm{E}| \times |\cm{V}|$ matrix with entries $R(e, v) = \gamma_e(v)$. The \textit{hyperedge-weight} matrix $W$ is a $|\cm{V}| \times |\cm{E}|$ matrix with entries $W(v, e) = \omega(e)$ if $v \in e$, and $W(v, e) = 0$ otherwise. The \textit{vertex-degree matrix} $D_{\cm{V}}$ is a $|\cm{V}| \times |\cm{V}|$ diagonal matrix with entries $D_{\cm{V}}=d(v)$. The \textit{hyperedge-degree matrix} $D_{\cm{E}}$ is a $|\cm{E}| \times |\cm{E}|$ diagonal matrix with entries $D_{\cm{E}}(e, e) = \delta(e)$.
\end{definition}
Since we are dealing with clustering, we make the following assumption similar to the graph setting.
\begin{assumption}
\label{assumption: hypergraph connectivity}
    Without loss of generality, we assume the hypergraph is connected. 
\end{assumption}

\subsubsection{Random walks on EDVW Hypergraphs}
Previous work \citep{DBLP:conf/icml/ChitraR19} introduces random walks on EDVW hypergraphs. We use the same notation as graph random walks, as both are fundamentally Markov processes.

\begin{definition}
\label{df: transition matrix}
 \citep{DBLP:conf/icml/ChitraR19} (Hypergraph random walk).
A \textit{random walk} on a hypergraph with edge-dependent vertex weights $\cm{H} = (\cm{V}, \cm{E}, \omega, \gamma)$ is a Markov Chain on $\cm{V}$ with transition probabilities 
\begin{equation}
    P_{u, v} = \sum_{e \in E(u)} \frac{\omega(e)}{d(u)} \frac{\gamma_e(v)}{\delta(e)}
\end{equation}
$P$ can be written in matrix form as $P = D_\cm{V}^{-1}WD_\cm{E}^{-1}R$ \citep{DBLP:conf/icml/ChitraR19} and it has row sum of 1. 
\end{definition}

\begin{definition}\label{df: stationary distribution}
(Stationary distribution of hypergraph random walk).
The \textit{stationary distribution} of the random walk with transition matrix $P$ is a $ 1 \times|\cm{V}|$ row vector $\phi$ such that
\begin{equation}
    \phi P = \phi;~ \phi(u) > 0 \,\, \forall u\in \cm{V};~ \sum_{u\in{\cm{V}}}\phi(u) = 1
\end{equation}
\end{definition}
The existence of $\phi$ for connected hypergraphs has been proved in \citep{DBLP:conf/icml/ChitraR19}. From $\phi$, we further define the \textit{stationary distribution matrix} to be a $|\cm{V}| \times |\cm{V}|$ diagonal matrix with entries $\Pi_{i, i} = \phi(i)$.

\section{Local Clustering}
\label{section: local}
\subsection{Stationary-Distributed Volumes}
From Definition \ref{df: p(S)} to Definition \ref{df: volume of a set}, in the context of EDVW hypergraph, following the definitions on graphs, we re-define the volume of boundaries and vertex sets, which are crucial for local clustering. We show that our definitions have properties that are consistent with those of graphs. Given a vertex set $\cm{S} \subseteq \cm{V}$, we use $\bar{\cm{S}}$ to denote its \textit{complementary set}, where $\cm{S} \cup \bar{\cm{S}} = \cm{V}$ and $\cm{S} \cap \bar{\cm{S}} = \emptyset$. 

\begin{definition}
\label{df: p(S)}
(Probability of a set).
    For a distribution $p$ on the vertices such that $\forall v \in \cm{V}, p(v) \geq 0$ and $\sum_{v \in \cm{V}}p(v) = 1$, we denote
\begin{equation}
    p(S) = \sum_{x \in S} p(x), \forall S \subseteq V
\end{equation}
\end{definition}
Equivalently, we can regard $p$ as a $1 \times |\cm{V}|$ vector with $p_i = p(i)$. From this definition, $\phi(\cm{S}) + \phi(\bar{\cm{S}}) = 1$. By definition of random walk and stationary distribution,
\begin{equation}
\begin{split}
    \phi&(S) = (\phi P)(S)\\
            &= {\phi(S)} - \underbrace{\sum_{u\in \cm{S}, v\in \bar{\cm{S}}} \phi(u)P_{u, v}}_{\text{walking out of }S} + \underbrace{\sum_{u\in \bar{\cm{S}}, v\in \cm{S}} \phi(u)P_{u, v}}_{\text{walking into }S}
\end{split}
\end{equation}
Therefore, we have the following Theorem~\ref{thm: equality of in and out} that, in the stationary state, for any set $\cm{S}$, the probability of walking into $\cm{S}$ or out of $\cm{S}$ are the same (rigorous proof in Lemma \ref{lem: partial_in = partial_out = partialS}).

\begin{theorem}
\label{thm: equality of in and out}
    Let $P$ be the transition matrix and $\phi$ be the stationary distribution of EDVW hypergraph $\cm{H} = (\cm{V}, \cm{E}, \omega, \gamma)$. Then, for any vertex set $\cm{S} \subseteq \cm{V}$,
\begin{equation}
    \sum_{u\in \cm{S}, v\in \bar{\cm{S}}} \phi(u)P_{u, v} = \sum_{u\in \bar{\cm{S}}, v\in \cm{S}} \phi(u)P_{u, v}
\end{equation}
\end{theorem}
For unweighted and undirected graphs, the volume of the \textit{boundary/cut} of a partition is defined as $|\partial\cm{S}| = |\{\{x, y\} \in E | x\in \cm{S}, y \in \bar{\cm{S}}\}|$. The intuition behind this definition is the symmetric property $|\partial\cm{S}| = |\partial\bar{\cm{S}}|$. Theorem \ref{thm: equality of in and out} also describes such a property, which can be extended to the hypergraph boundary.

\begin{definition}
\label{df: volume of boundary}
(Volume of hypergraph boundary).
    We define the \textit{volume of the hypergraph boundary}, i.e., cut between $\cm{S}$ and $\bar{\cm{S}}$, by
\begin{equation}
    |\partial S| = \sum_{u\in \cm{S}, v\in \bar{\cm{S}}} \phi(u)P_{u, v} = \sum_{u\in \bar{\cm{S}}, v\in \cm{S}} \phi(u)P_{u, v}
\end{equation}
Furthermore, $0 \leq |\partial \cm{S}| = |\partial\bar{\cm{S}}| \leq \sum_{u\in \cm{S}} \phi(u) \leq 1$.
\end{definition}

For unweighted, undirected graphs, the \textit{volume of a vertex set} $\cm{S}$ is defined as the degree sum of the vertices in $\cm{S}$. With $\phi(u) = \sum_{u\in \cm{S}, v\in \cm{V}} \phi(u)P_{u, v}$, $\phi(u)$ itself is already a sum of the transition probabilities and can be an analogy to vertex degree. We extend this observation to the following definition.

\begin{definition}
\label{df: volume of a set}
(Volume of hypergraph vertex set).
    We define the \textit{volume of a vertex set} $\cm{S} \subseteq \cm{V}$ in hypergraph $\cm{H}$ by
\begin{equation}
    vol(S) = \sum_{u\in \cm{S}} \phi(u) \in [0, 1]
\end{equation}
\end{definition}
Furthermore, $vol(\emptyset) = 0$, $vol(\cm{V}) = 1$, and $vol(\cm{S}) + vol(\bar{\cm{S}}) = 1$. Definition \ref{df: volume of boundary} and Definition \ref{df: volume of a set} will serve as the basis of our unified formulation. By these two definitions, we also have $|\partial \cm{S}| \leq vol(\cm{S})$, which is consistent with those on unweighted and undirected graphs.



\subsection{Conductance Measures as Markov Chains}
The conductance measures the proportion of "broken" edges relative to the total importance of the two clusters. It quantifies the quality of a cluster by balancing the separation between clusters and their internal connectivity. 
Following the previous work \citep{DBLP:journals/im/AndersenCL08, DBLP:conf/focs/AndersenCL06}, we present a generalized definition of conductance for graphs that may be edge-weighted, node-weighted, directed, and self-looped. This extends the original definition in \citep{DBLP:journals/im/AndersenCL08} to accommodate more complex graph structures.
\begin{definition}
\label{df: general graph conductance}
(General graph conductance).
    The \textit{conductance} of a cluster $\cm{S}$ on $\cm{G}$ with transition matrix $P$ and stationary distribution $\phi$ is,
\begin{equation}
\begin{split}
    \Phi_{\cm{G}}(\cm{S}) &= \frac{\sum_{u \in \cm{S}, v \in \cm{\bar{\cm{S}}}}\phi(u)P_{u,v}}{\min(\sum_{v \in \cm{S}}\phi(u), 1-\sum_{u \in \cm{S}}\phi(u))}
\end{split}
\end{equation}
\end{definition}

In this work, we follow the conventional formulation $\Phi(\cm{S}) = \frac{|\partial\cm{S}|}{\min(vol(\cm{S}), vol(\bar{\cm{S}}))}$ for conductance but applied to hypergraphs in EDVW formatting for the first time.

\begin{definition}
\label{df: conductance}
(Hypergraph conductance).
    The \textit{conductance} of a cluster $\cm{S}$ on $\cm{H}$ with transition matrix $P$ and stationary distribution $\phi$ is,
\begin{equation}
\resizebox{0.85\hsize}{!}{$
\begin{aligned}
    \Phi_{\cm{H}}(\cm{S}) &= \frac{|\partial\cm{S}|}{\min(vol(\cm{S}), vol(\bar{\cm{S}}))}
                = \frac{|\partial\cm{S}|}{\min(vol(\cm{S}), 1 - vol(\cm{S}))}\\
                &= \frac{\sum_{u \in \cm{S}, v \in \cm{\bar{\cm{S}}}}\phi(u)P_{u,v}}{\min(\sum_{v \in \cm{S}}\phi(u), 1-\sum_{u \in \cm{S}}\phi(u))}
\end{aligned}
$}
\end{equation}

\end{definition}

\begin{theorem}
    For any vertex set $\cm{S} \subseteq \cm{V}$, our hypergraph conductance $\Phi_{\cm{H}}(\cm{S}) \in [0, 1]$, which is consistent with graph conductance. (Proof in Appendix \ref{pf: Phi(S) bound})
\label{thm: Phi(S) bound}
\end{theorem}

Notably, the conductance of general graphs and hypergraphs shares the same formulation on their equivalent Markov chains. All later proofs leveraging conductance require no additional properties beyond this. Therefore, we unify the notation and use $\Phi$ to represent both $\Phi_{\cm{G}}$ (for graphs) and $\Phi_{\cm{H}}$ (for hypergraphs).

\subsection{GeneralACL and HyperACL Algorithms based on Personalized PageRank}

Given a starting vertex set, finding optimal local clusters in terms of minimal conductance is an NP-complete problem \citep{DBLP:conf/sofsem/SimaS06}. GeneralACL and HyperACL rank the vertex candidacy heuristically using seeded/personalized random walks. 

\begin{definition}
(Lazy PPR vector)
The \textit{lazy personalized PageRank vector} $pr(\alpha, s)$ on $\cm{H}$ satisfies 
\begin{equation}
    pr(\alpha, s) = \alpha s + (1-\alpha) pr(\alpha, s) M
\end{equation}
where $M = \frac{1}{2}(I + P)$, $s$ is the \textit{stochastic vector} of random walks, $\alpha$ is the restart probability. 
\label{df: pagerank}
\end{definition}
In fact, a lazy random walk is equivalent to a standard random walk by a switch of $\alpha$, as Lemma \ref{lem: equivalent of lazy and standard} states.

\begin{algorithm}[t]
   \caption{GeneralACL and HyperACL}
   \label{alg: local graph}
\begin{algorithmic}[1]
    \REQUIRE Graph $\cm{G} = (\cm{V}, W, \varphi)$ \textit{or} EDVW hypergraph $\cm{H} = (\cm{V}, \cm{E}, \omega, \gamma)$; a set of starting vertices $\cm{S}$ for local clustering
    \ENSURE a cluster of vertices
   \STATE For \underline{graph} setting, compute transition matrix from assumption \ref{assumption: well-defined transition matrix}; for \underline{hypergraph} setting, Compute $R, W, D_\cm{V}, D_\cm{E}$ from Definition \ref{df: R, W, Dv, De}. then compute transition matrix $P$ by Definition \ref{df: transition matrix}.
   \STATE Compute stationary distribution by power iteration.
   \STATE Compute PageRank vector as in Theorem \ref{thm: conductance bound}.
   \STATE Compute sweep sets by Definition \ref{df: sweep}.
   \STATE Compute which sweep set has the smallest conductance with early stop and return this sweep set.
\end{algorithmic}
\end{algorithm}


\begin{definition}
    The \textit{indicator function} $\chi_v$ is \\ $\chi_v(x) = \begin{cases}
        1 & x= v\\
        0 & otherwise\\
    \end{cases}$  
and we denote $pr(\alpha, \chi_v)$ to be the \textit{lazy single-source PageRank vector} of $v$.
\label{df: indicator function and sspr}
\end{definition}


Sweep sets serve as an effective manner for obtaining the local clustering as follows.

\begin{definition}
\citep{DBLP:conf/focs/AndersenCL06} (Sweep sets/cuts). For a distribution $p$ on the vertices such that $\forall v \in \cm{V}, p(v) \geq 0$; $\sum_{v \in \cm{V}}p(v) = 1$, let $N_p = |Supp(p)|$ be the support size (a.k.a., number of vertices $v$ with $p(v) \neq 0$), and let $v_1, v_2, ..., v_{N_p}$ be an ordering of vertices such that $\frac{p(v_i)}{\phi(v_i)} \geq \frac{p(v_{i+1})}{\phi(v_{i+1})}$. From this ordering we can obtain $N_p$ vertex sets $\cm{S}_1^p, \cm{S}_2^p, ...,  \cm{S}_{N_p}^p$ such that $\cm{S}_j^p= \{v_1, v_2, ..., v_j\}, \, \forall j \in [1, N_p]$. We call the ordered sets $\cm{S}_1^p, \cm{S}_2^p, ...,  \cm{S}_{N_p}^p$ \textit{sweep sets} or \textit{sweep cuts}. Each $\cm{S}_j^p$ is called a \textit{sweep set/cut}.
\label{df: sweep}
\end{definition}

Directly from the above definition, we have $p(\cm{S}_j^p) \geq 0$, $p(\cm{S}_j^p) < p(\cm{S}_{j+1}^p)$, $p(\cm{S}_{N_p}^p) = 1$. According to the monotonicity of $\frac{p(v_i)}{\phi(v_i)}$, we have $\frac{p(\cm{S}_j^p)}{vol(\cm{S}_j^p)} \geq \frac{p(\cm{S}_{j+1}^p)}{vol(\cm{S}_{j+1}^p)} \,\, \forall j \in [1, N_p - 1]$; and $\frac{p(\cm{S}_j^p)}{vol(\cm{S}_j^p)} \geq \frac{p(\cm{S})}{vol(\cm{S})} \,\, \forall \cm{S} \subseteq |V| \, s.t. \,|\cm{S}| = |\cm{S}_j^p|$. 

\begin{definition}
\label{df: optimal conductance of a distribution}
\citep{DBLP:conf/focs/AndersenCL06} (Optimal conductance of a distribution)
    We denote the \textit{optimal conductance of a distribution} to be the smallest conductance of any of its sweep sets.
\begin{equation}
    \Phi(p) = \min_{j \in [1, N_p]} \Phi(\cm{S}_j^p)
\end{equation}
which can be found by sorting $\frac{p(v_i)}{\phi(v_i)}$ and computing the conductance of each sweep set.
\end{definition}

\textbf{Our GeneralACL and HyperACL Algorithms.} Given seed vertices, we calculate the random walk probability starting from these seed vertices, a.k.a. PageRank vector, and the corresponding sweep. The vertices that are ranked in the front of the sweep have high seeded-random-walk distributions, weighted by the stationary random walk distributions. Heuristically, this means the vertices that are ranked in the front are relatively more closely connected to the seed vertices. Formally, Algorithm \ref{alg: local graph} The ultimate goal of the later sections is to prove \textbf{Theorem \ref{thm: conductance bound}}, which shows that GeneralACL and HyperACL, leveraging the Markov chains induced by random walks on either general graph or hypergraph, in many situations, find a quadratically optimal local cluster in terms of conductance.

It is worth noting that, in practice, the most time-consuming step is checking the conductance of every sweep set. 
We adopt an early-stop mechanism here to accelerate the computation: when sweeping over the candidate local clusters, if the conductance does not break the minimum within a manually set hyperparameter, then we assume we have obtained a local cluster that is good enough.  

\begin{theorem}[Worst-case Time Complexity without Early Stop]
Let $m$ denote the number of hyperedge--vertex incidences, 
$|\cm{V}|$ the number of vertices, and $k$ the size of the returned local cluster.
HyperACL has worst-case time complexity
$O\!\left(m^2 + |\cm{V}|^2 + k^2|\cm{V}|\right)$.
GeneralACL has worst-case time complexity
$
O\!\left(C + |\cm{V}|^2 + k^2|\cm{V}|\right),
$
where $C$ is the cost of computing the transition matrix. (Proof in Appendix \ref{sec: time complexity hyper l})
\end{theorem}

\begin{theorem}[Worst-case Space Complexity without Early Stop]
Under the same notation, the worst-case space complexity of
HyperACL is
$
O\!\left(m + |\cm{V}|^2\right),
$
and GeneralACL is
$
O\!\left(C + |\cm{V}|^2\right),
$
where $C$ is the space required for computing the transition matrix. (Proof in Appendix \ref{sec: space complexity hyper l})
\end{theorem}

Although the worst-case bounds are quadratic, such costs are typical for exact PageRank-based clustering.
In practice, scalability is often improved via sparse/localized computation,
approximate PageRank (e.g., push or truncated diffusion), parallel or distributed implementations,
yielding near-linear complexity on many real-world graphs.
In this work, we focus on the exact formulation to obtain clean theoretical guarantees,
while our early-stop mechanism provides additional practical speedups.

\section{Extending Andersen-Chung-Lang Algorithm}
\label{sec: local_proof}

\subsection{Lovász-Simonovits Curve}
\label{sec: LS curve}

Lovász-Simonovits Curve (LSC or L-S Curve) defines a piecewise function. Unlike the original proof on unweighted undirected graphs, we cannot simply use the integer node degrees to define the volume of a vertex set. In our proof, the support of Lovász-Simonovits Curve is [0, 1] instead of [0, $2|\cm{E}|$]..

\begin{definition}
    Given a hypergraph $\cm{H} = (\cm{V}, \cm{E}, \omega, \gamma)$, a distribution $p$ and their corresponding sweep sets $\cm{S}_1^p, \cm{S}_2^p, ...,  \cm{S}_{N_p}^p$, Lovász-Simonovits Curve defines a piecewise function $I_p: [0, 1] \rightarrow [0, 1]$ such that $\forall k\in [0, 1]$, using $p[k]$ for short,
\begin{equation}
\resizebox{1\hsize}{!}{$
\begin{aligned}
    & p[k] = I_p(k) \\
    & = \begin{cases}
        0 & k = 0\\
        p(\cm{S}_j^p) & if \,\, \exists j \,\, s.t. \,\, k=vol(\cm{S}_j^p)\\
        p(\cm{S}_j^p) + (k - vol(\cm{S}_j^p)) \frac{p(v_{j+1})}{\phi(v_{j+1})} & if \,\, \exists j \,\, s.t. \,\, vol(\cm{S}_j^p)<k<vol(\cm{S}_{j+1}^p)\\
        1 & vol(\cm{S}_{N_p}^p) < k \leq 1 \\
    \end{cases}
\end{aligned}
$}
\end{equation}
\label{df: LSC}
\end{definition}
\noindent $p[k]$ is continuous because
\begin{equation}
\resizebox{1\hsize}{!}{$
\begin{aligned}
    & \mathrm{when} \,\, k = vol(\cm{S}_j^p), p(\cm{S}_j^p) + (k - vol(\cm{S}_j^p)) \frac{p(v_{j+1})}{\phi(v_{j+1})} = p(\cm{S}_j^p) \\
    & \mathrm{when} \,\, k = vol(\cm{S}_{j+1}^p), p(\cm{S}_j^p) + (k - vol(\cm{S}_j^p)) \frac{p(v_{j+1})}{\phi(v_{j+1})} = vol(\cm{S}_{j+1}^p) \\ & \qquad \qquad \qquad \qquad = p(\cm{S}_j^p) +  \phi(v_{j+1})\frac{p(v_{j+1})}{\phi(v_{j+1})} = p(\cm{S}_{j+1}^p) \\
    & \mathrm{when} \,\,  k = vol(\cm{S}_{N_p}^p), p(\cm{S}_{N_p}^p) = 1 \\
\end{aligned}
$}
\end{equation}

Furthermore, $p[k]$ is concave because $\frac{p(\cm{S}_j^p)}{vol(\cm{S}_j^p)} \geq \frac{p(\cm{S}_{j+1}^p)}{vol(\cm{S}_{j+1}^p)} \,\, \forall j \in [1, N_p - 1]$. 

The following definitions and lemmas about the Lovász-Simonovits Curve are important for later proofs. 

\begin{definition}
    For any node pair $u, v \in \cm{V}$ (it is possible that $u = v$), we denote $p(u, v) = \frac{p(u)}{\phi(u)} P_{u, v}$.
\label{df: p(u, v)}
\end{definition}

\begin{definition}
    For any edge set $\cm{A}$, we define notation $p(\cm{A})$ as $p(\cm{A}) = \sum_{(u, v) \in \cm{A}}\phi(u)p(u, v) = \sum_{(u, v) \in \cm{A}}p(u)P_{u, v}$.
\label{df: p(A)}
\end{definition}

\begin{definition}
    For any edge set $\cm{A}$, we define notation $|\cm{A}|$ as $|\cm{A}| = \sum_{(u, v) \in \cm{A}}\phi(u)P_{u, v}$.
\label{df: |A|}
\end{definition}

Note that in the above definitions, the edge set $\cm{A} \in \cm{V} \times \cm{V}$ is not a hyperedge set, but only contains node pairs. Also, for an edge set $\cm{A}$, $|\cm{A}|$ in this paper does not refer to the cardinal number of $\cm{A}$.

\subsection{Upper Bound of the Disagreement}
\label{sec: upper bound of disagreement}
In this part, we aim to set up an upper bound of $p[k] - k, \forall k\in [0, 1]$ (Theorem \ref{thm: either or}). For any vertex set, The following definition describes the edges going into it and the edges going outside of it.
\begin{definition}
    For any vertex set $\cm{S} \in \cm{V}$, we denote $in(\cm{S}) = \{(u, v)| u\in \cm{V}, v\in \cm{S}\}$; $out(S) = \{(u, v)| u\in \cm{S}, v\in \cm{V}\}$.
\label{df: in out}
\end{definition}

From this definition, we set up some lemmas (Lemma \ref{lem: pms in out} to Lemma \ref{lem: pvol pvol+-partial}) to prove Theorem \ref{thm: either or}.

\begin{lemma}
    For any distribution $p$ and any vertex set $\cm{S} \subseteq \cm{V}$, recall definition \ref{df: p(A)},
\begin{equation}
\resizebox{0.85\hsize}{!}{$
\begin{aligned}
    (pM)(S) &= \frac{1}{2}p(in(\cm{S})) + \frac{1}{2}p(out(\cm{S})) \\
    &= \frac{1}{2}p(in(\cm{S}) \cup out(\cm{S})) + \frac{1}{2}p(in(\cm{S})\cap p(out(\cm{S}))
\end{aligned}
$}
\end{equation}
\label{lem: pms in out}
(Proof in Appendix \ref{pf: pms in out})
\end{lemma}

\begin{lemma}
    Recall the definition of $|\cm{A}|$ from definition \ref{df: |A|}. For any two sets of node pairs $\cm{A}, \cm{B}$,
\begin{equation}
    |\cm{A}| + |\cm{B}| = |\cm{A} \cup \cm{B}| + |\cm{A} \cap \cm{B}|
\end{equation}
\label{lem: ababab}
(Proof in Appendix \ref{pf: ababab})
\end{lemma}

\begin{lemma}
    Recall the definition of $|\cm{A}|$ from definition \ref{df: |A|}, $in(\cm{S})$ and $out(\cm{S})$ from definition \ref{df: in out}, and $vol(\cm{S})$ from definition \ref{df: volume of a set}. For any vertex set $\cm{S} \subseteq \cm{V}$,
\begin{equation}
    |in(\cm{S})| + |out(\cm{S})| = 2 vol(\cm{S})
\end{equation}
More specifically, 
\begin{equation}
    |in(\cm{S})| = |out(\cm{S})| = vol(\cm{S})
\end{equation}
\label{lem: in out vol}
(Proof in Appendix \ref{pf: in out vol})
\end{lemma}

\begin{lemma}
    Recall the definition of $|\cm{A}|$ from definition \ref{df: |A|}, and $|\partial \cm{S}|$ from definition \ref{df: volume of boundary}. For any vertex set $\cm{S} \subseteq \cm{V}$,
\begin{equation}
    |in(\cm{S})\cup out(\cm{S})| - |in(\cm{S})\cap out(\cm{S})| = 2|\partial \cm{S}|
\end{equation}
\label{lem: in out partial}
(Proof in Appendix \ref{pf: in out partial})
\end{lemma}

\begin{lemma}
    For any vertex set $\cm{S} \subseteq \cm{V}$,
\begin{equation}
\begin{split}
    |in(\cm{S})\cup out(\cm{S})| &= vol(\cm{S}) + |\partial \cm{S}| \\
    |in(\cm{S})\cap out(\cm{S})| &= vol(\cm{S}) - |\partial \cm{S}| \\
\end{split}
\end{equation}
\label{lem: in out vol and partial}
(Proof in Appendix \ref{pf: in out vol and partial})
\end{lemma}

\begin{lemma}
    If $p = pr(\alpha, s)$ is a lazy PageRank vector, then $\forall \cm{S} \in \cm{V}$,
\begin{equation}
\resizebox{1\hsize}{!}{$
    p(\cm{S}) = \alpha s(\cm{S}) + (1-\alpha) \frac{1}{2}(p(in(\cm{S})\cup out(\cm{S})) + p(in(\cm{S})\cap out(\cm{S})))
$}
\end{equation}
    Furthermore, recall the definition of $p[\cdot]$ from definition \ref{df: LSC}, for each $j \in [1, N_p]$, 
\begin{equation}
\resizebox{0.9\hsize}{!}{$
\begin{aligned}
    & p[vol(\cm{S}_j^p)] \leq \alpha s[vol(\cm{S}_j^p)] \\
    & + (1-\alpha) \frac{1}{2}(p[vol(\cm{S}_j^p) + |\partial \cm{S}_j^p|] + p[vol(\cm{S}_j^p) - |\partial \cm{S}_j^p|])
\end{aligned}
$}
\end{equation}
\label{lem: pvol pvol+-partial}
(Proof in Appendix \ref{pf: pvol pvol+-partial})
\end{lemma}

\begin{theorem}
    Let $\sigma$ $\gamma$ and $\theta$ be any constants such that $\sigma \in [0, 1], \gamma \in [0, 1]$, $\theta \in [0, \frac{1}{2}]$.
    Let $p = pr(\alpha, s)$ be a lazy PageRank vector, then
    \textbf{either} the following bound holds for any integer $t$ and any $k \in [\theta, 1-\theta] \cup \{0, 1\}$,
\begin{equation}
    p[k] - k \leq \gamma + \alpha t + \sqrt{\frac{1}{\theta}}\sqrt{\min(k, 1-k)} (1 - \frac{\sigma^2}{8})^t
\end{equation}
\textbf{or} there exists a sweep cut $\cm{S}_j^p, j \in [1, N_p]$, with the following properties,
\begin{equation}
\begin{split}
    & \mathrm{Property}\, 1.\, \Phi(\cm{S}_j^p) < \sigma\\
    & \mathrm{Property}\, 2.\, \exists t \in \mathbb{N}_+, p(\cm{S}_j^p) - vol(\cm{S}_j^p) > \gamma + \alpha t + \\
    & \qquad \sqrt{\frac{1}{\theta}}\sqrt{\min(vol(\cm{S}_j^p), 1-vol(\cm{S}_j^p))} (1 - \frac{\sigma^2}{8})^t
\end{split}
\end{equation}
\label{thm: either or}
(Proof in Appendix \ref{pf: either or})
\end{theorem}

\begin{figure*}[t]
\centering
\subfigure[DBLP-ML $\Phi$]{
\includegraphics[width=0.22\textwidth]{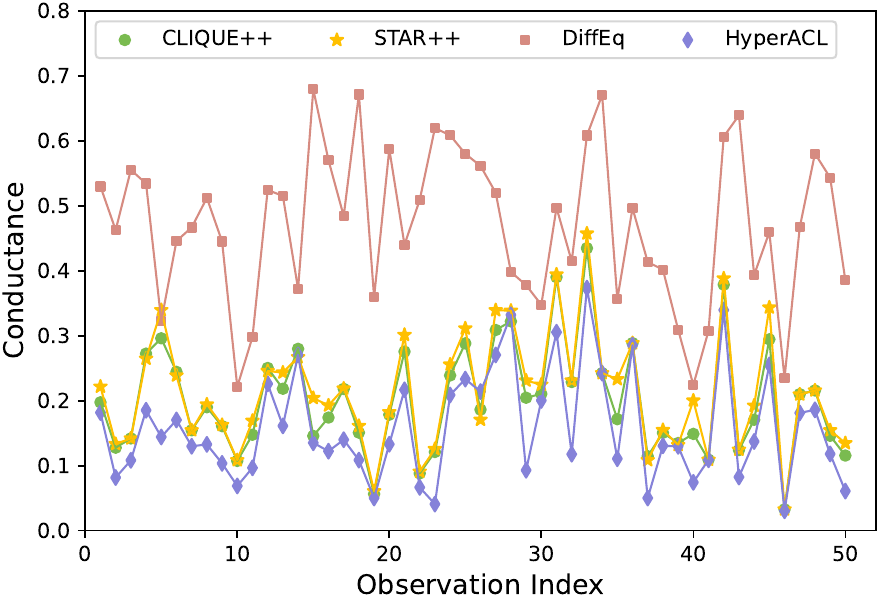}
\label{Fig2-1}
}
\subfigure[DBLP-CV $\Phi$]{
\includegraphics[width=0.22\textwidth]{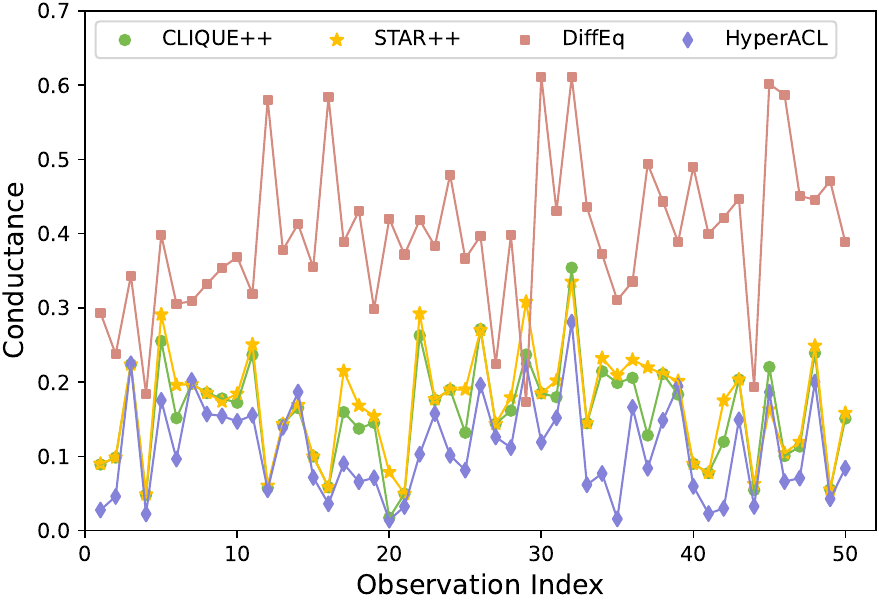}
\label{Fig2-2}
}
\subfigure[DBLP-NLP $\Phi$]{
\includegraphics[width=0.22\textwidth]{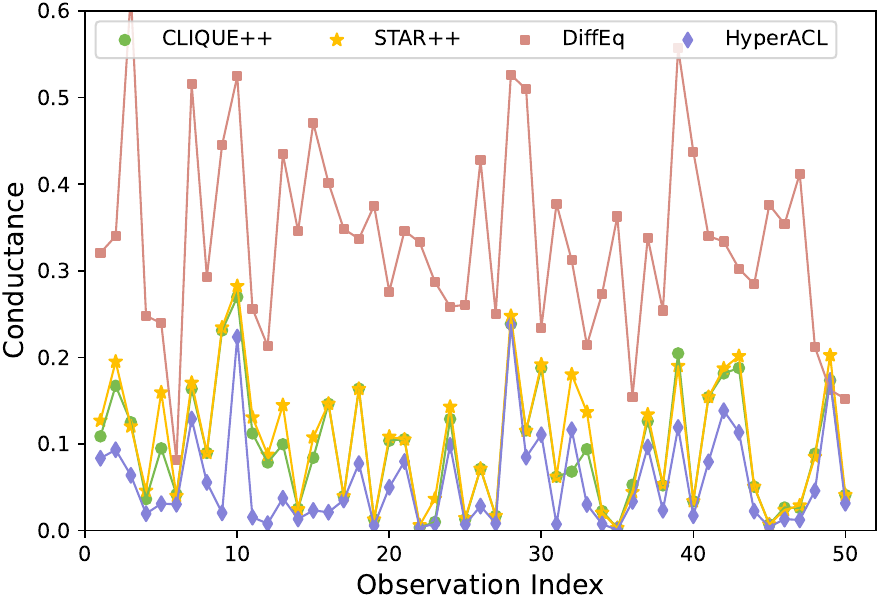}
\label{Fig2-3}
}
\subfigure[DBLP-IR $\Phi$]{
\includegraphics[width=0.22\textwidth]{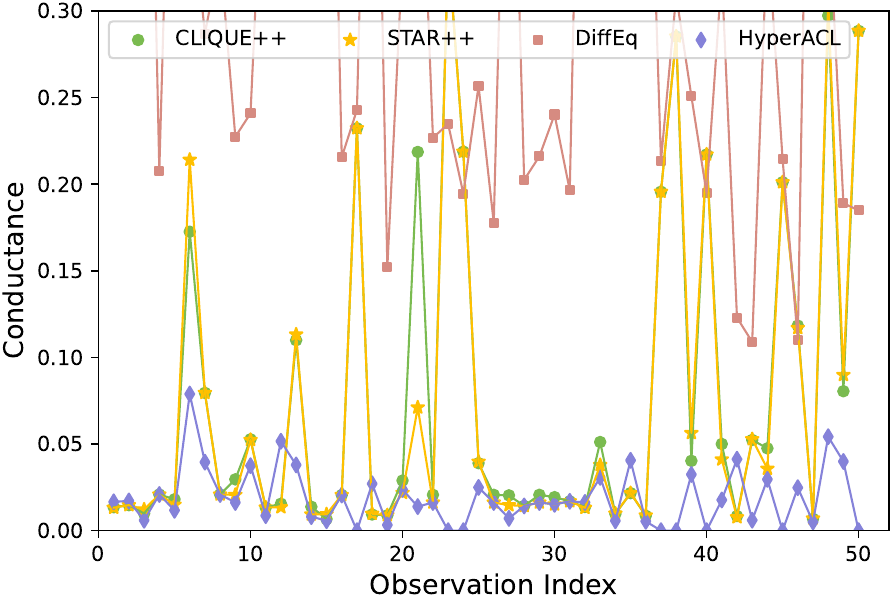}
\label{Fig2-4}
}
\subfigure[DBLP-ML F1]{
\includegraphics[width=0.22\textwidth]{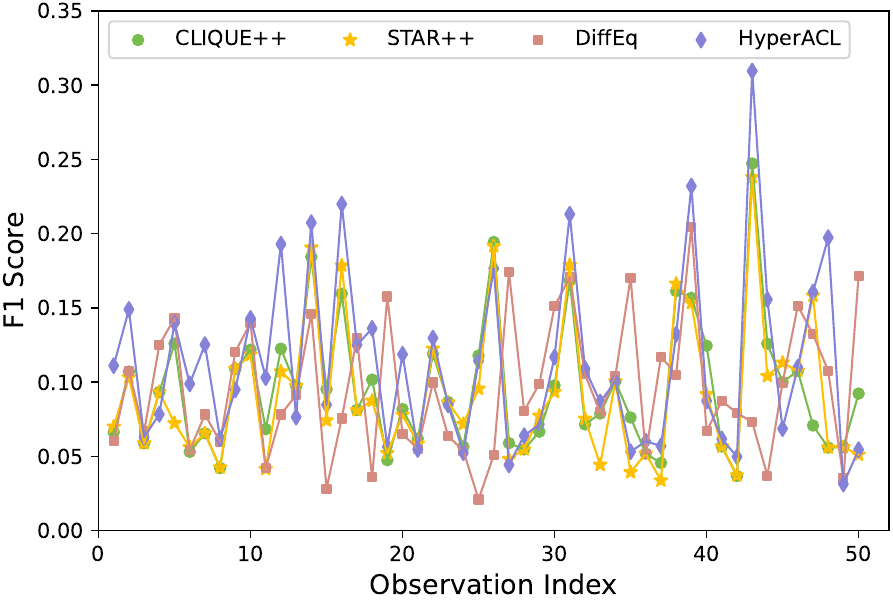}
\label{Fig2-5}
}
\subfigure[DBLP-CV F1]{
\includegraphics[width=0.22\textwidth]{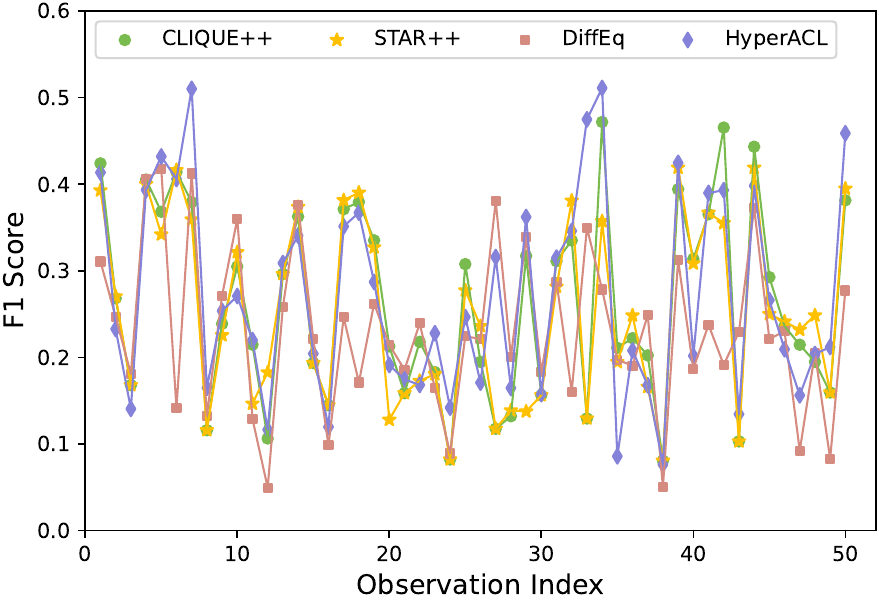}
\label{Fig2-6}
}
\subfigure[DBLP-NLP F1]{
\includegraphics[width=0.22\textwidth]{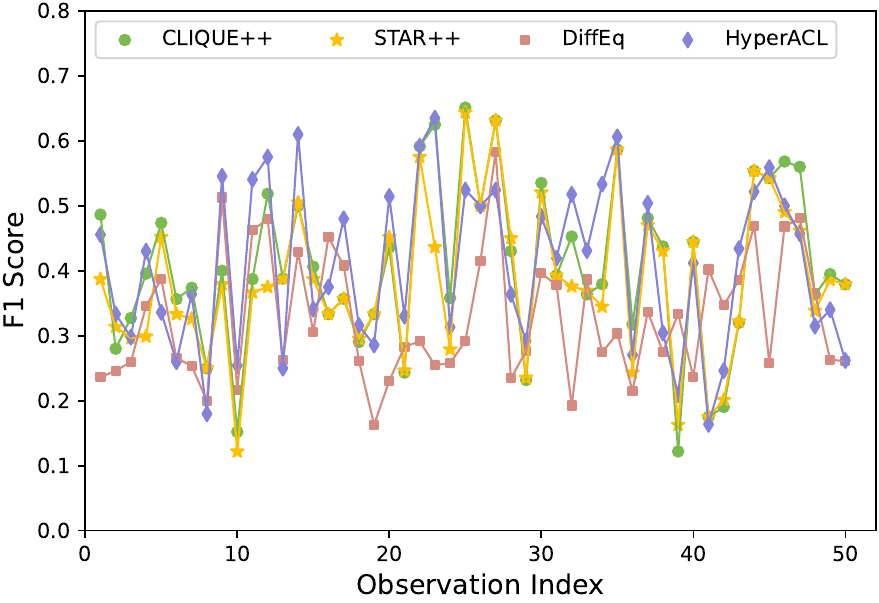}
\label{Fig2-7}
}
\subfigure[DBLP-IR F1]{
\includegraphics[width=0.22\textwidth]{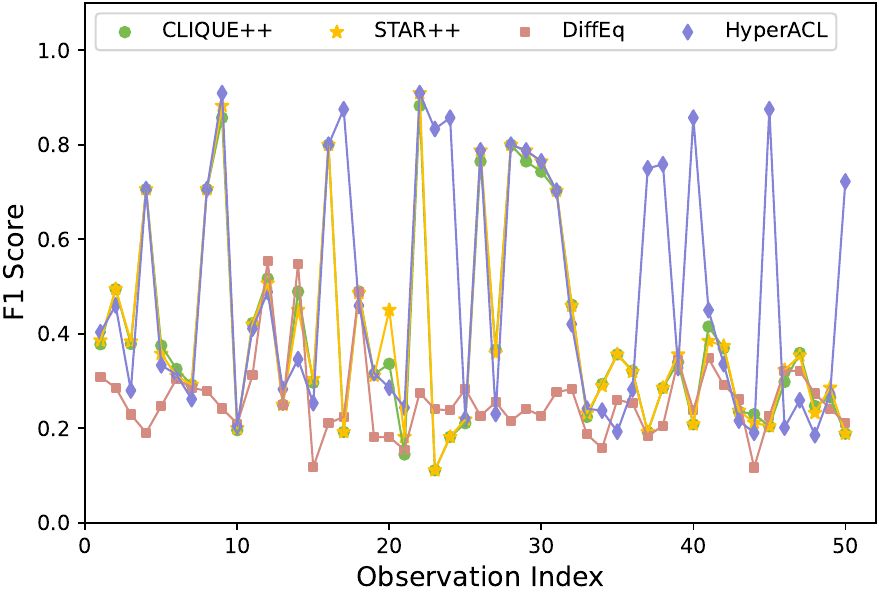}
\label{Fig2-8}
}
\caption{Conductance $\Phi$ ($\downarrow$) and F1 ($\uparrow$) Comparison on the Local Clustering Task.}
\label{fig: local results}
\end{figure*}

\subsection{Solve the Conductance Given $p[k] - k$ Lower Bound}
\label{sec: solve conductance given lower bound}
From Theorem \ref{thm: either or}, since the right-hand side on the ``either" side decreases when $\sigma$ increases, if $p[k] - k$ has another lower bound, then we can further upper-bound $\sigma = \Phi(pr(\alpha, s))$.

\begin{theorem}
    For a lazy PageRank vector $pr(\alpha, s)$ and any constant $z \in [0, \frac{1}{e}]$, if there exists a vertex set $\cm{S} \subseteq \cm{V}$ that satisfies

\begin{equation}
    \delta = pr(\alpha, s)(\cm{S}) - vol(\cm{S}) > z
\end{equation}
then,
\begin{equation}
    \Phi(pr(\alpha, s)) < \sqrt{\frac{9\alpha \ln\frac{1}{z}}{\delta - z}}
\end{equation}
\label{thm: conductance_satisfying_shrunk}
(Proof in Appendix \ref{pf: conductance_satisfying_shrunk})
\end{theorem}

\subsection{Lower Bound of the Disagreement}
\label{sec: lower bound of disagreement}
In this part, we find a $z$ value in Theorem \ref{thm: conductance_satisfying_shrunk} to bound $\Phi(pr(\alpha, s))$ in the next subsection.

\begin{definition}
    For any vertex set $\cm{S} \subseteq \cm{V}$, we denote $\partial_{in}(\cm{S}) = \{(u, v)| u\in \bar{\cm{S}}, v\in \cm{S}\}$; $\partial_{out}(S) = \{(u, v)| u\in \cm{S}, v\in \bar{\cm{S}}\}$.
    Then, by Definition \ref{df: |A|}, we have $|\partial_{in}(\cm{S})| = \sum_{u\in \bar{\cm{S}}, v\in \cm{S}} \phi(u) P_{u, v}, |\partial_{out}(\cm{S})| = \sum_{u\in \cm{S}, v\in \bar{\cm{S}}} \phi(u) P_{u, v}$.
\label{df: partial in out}
\end{definition}

Here, we give another proof for Theorem \ref{thm: equality of in and out} which shows $|\partial_{in}(\cm{S})| = |\partial_{out}(\cm{S})| = |\partial \cm{S}|$.

\begin{lemma}
    For any vertex set, referring to definition \ref{df: partial in out}, $|\partial_{in}(\cm{S})| = |\partial_{out}(\cm{S})| = |\partial \cm{S}|$.
\label{lem: partial_in = partial_out = partialS}
(Proof in \ref{pf: partial_in = partial_out = partialS})
\end{lemma}

\begin{definition}
    For any vertex set $\cm{C} \subseteq \cm{V}$, we define its \textit{PageRank Starting Distribution}
\begin{equation}
    \psi_\cm{C}(v) = \begin{cases}
        \frac{\phi(v)}{vol(\cm{C})} & ,\,\, v \in \cm{C}\\
        \,\, 0                           & ,\,\, otherwise\\
    \end{cases}
\end{equation}
\label{df: psi}
\end{definition}

\begin{theorem}
    For any vertex set $\cm{C} \subseteq \cm{V}$, recall the definition of lazy PageRank vector in Definition \ref{df: pagerank} and Definition \ref{df: p(S)},
\begin{equation}
    pr(\alpha, \psi_{\cm{C}})(\bar{\cm{C}}) \leq \frac{\Phi(\cm{C})}{2\alpha}
\end{equation}
\label{thm: psi}
(Proof in Appendix \ref{pf: psi})
\end{theorem}


\begin{figure*}[t]
\centering
\subfigure[DBLP-ML Running Time]{
\includegraphics[width=0.22\textwidth]{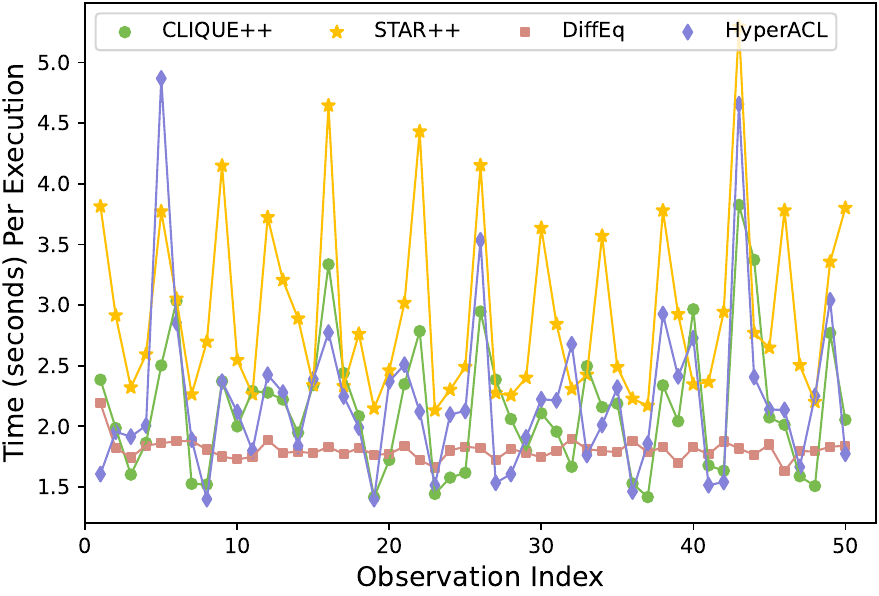}
\label{Fig3-1}
}
\subfigure[DBLP-CV Running Time]{
\includegraphics[width=0.22\textwidth]{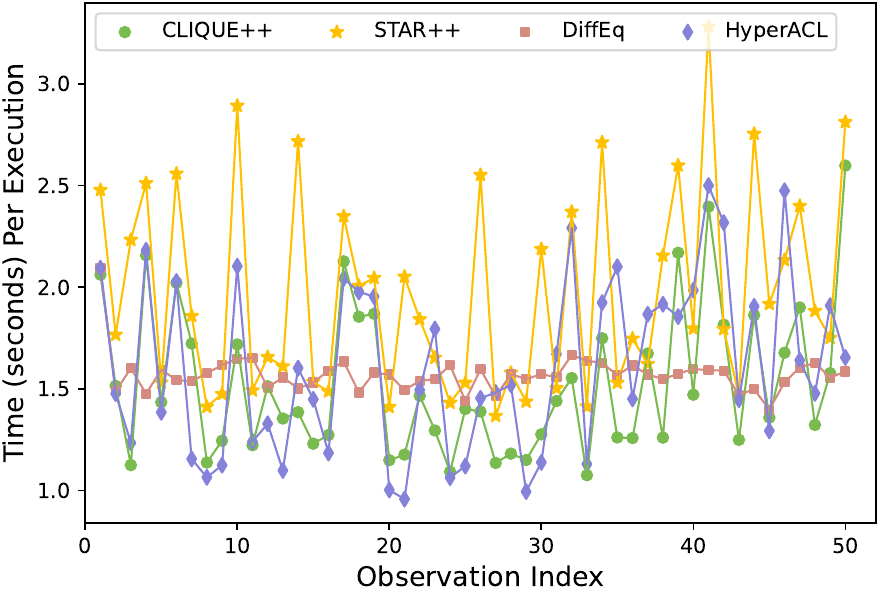}
\label{Fig3-2}
}
\subfigure[DBLP-NLP Running Time]{
\includegraphics[width=0.22\textwidth]{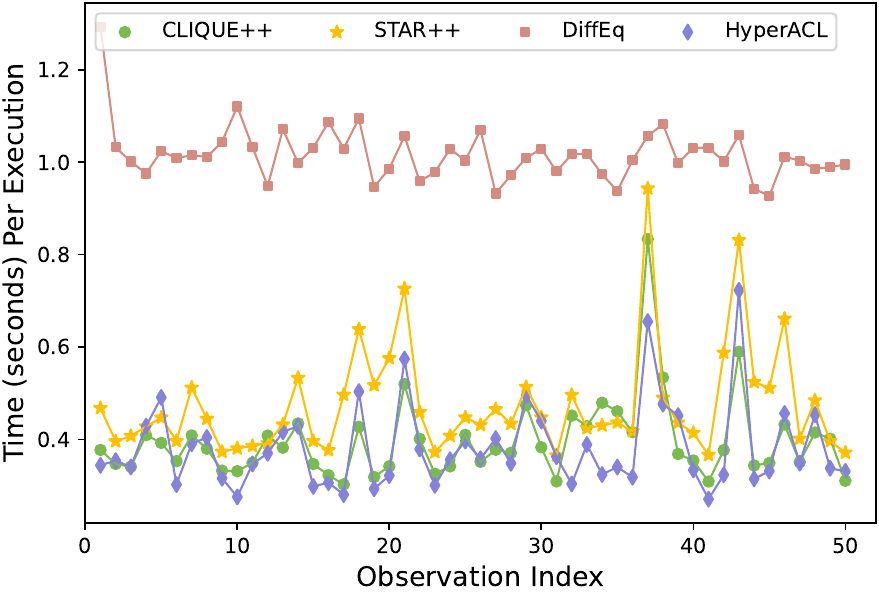}
\label{Fig3-5}
}
\subfigure[DBLP-CV Running Time]{
\includegraphics[width=0.22\textwidth]{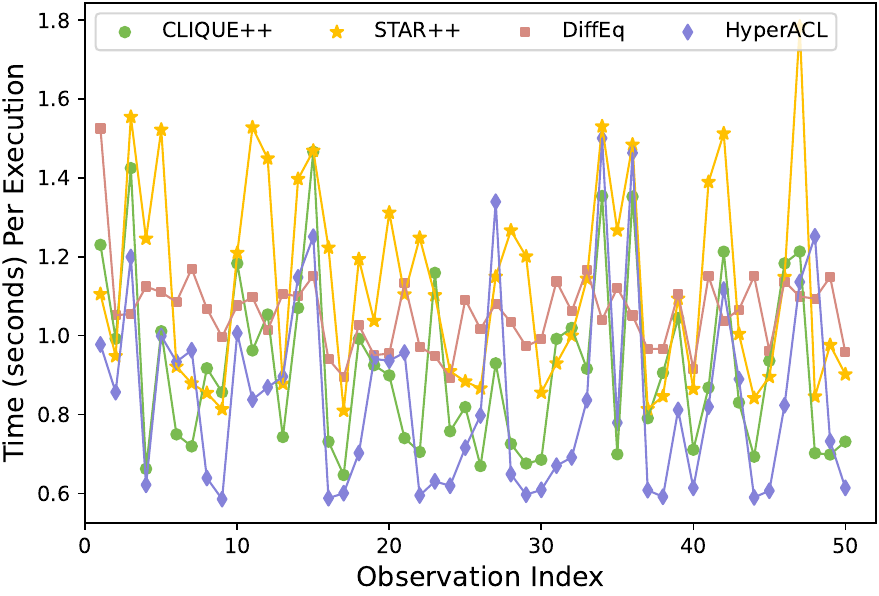}
\label{Fig3-6}
}
\caption{Running Time ($\downarrow$) Comparison on the Local Clustering Task.}
\label{fig: local times}
\end{figure*}

\subsection{Quadratic Optimality}
\label{sec: quadratic optimality}
\begin{theorem}
    For any vertex set $\cm{C} \subseteq \cm{V}$, sample a subset $\cm{D} \subseteq \cm{C}$ uniformly randomly. Then, with probability at least $\frac{1}{2}$,
\begin{equation}
    pr(\alpha, \psi_{\cm{D}})(\bar{\cm{C}}) \leq pr(\alpha, \psi_{\cm{C}})(\bar{\cm{C}})
\end{equation}
Moreover, at least one of the following equations holds.
\begin{equation}
    pr(\alpha, \psi_\cm{D})(\bar{\cm{C}}) \leq pr(\alpha, \psi_{\cm{C}})(\bar{\cm{C}})
\end{equation}
\begin{equation}
    pr(\alpha, \psi_{\cm{C}\setminus\cm{D}})(\bar{\cm{C}}) \leq pr(\alpha, \psi_{\cm{C}})(\bar{\cm{C}})
\end{equation}

\label{thm: PhiD no greater than PhiC}
(Proof in Appendix \ref{pf: PhiD no greater than PhiC})
\end{theorem}

\begin{theorem}
\label{thm: conductance bound}
    For a vertex set $\cm{S} \subseteq \cm{V}$ that is uniformly randomly sampled from the subsets of $\cm{V}$, let $\cm{S}^*$ be the vertex set with optimal conductance among all the vertex sets containing $\cm{S}$. Let the \textit{PageRank Starting Distribution} be
\begin{equation}
    \psi_\cm{S}(v) = \begin{cases}
        \frac{\phi(v)}{vol(\cm{S})} & ,\,\, v \in \cm{S}\\
        \,\, 0                           & ,\,\, otherwise\\
    \end{cases}
\end{equation}
and let $\alpha = \Phi(\cm{S}^*) \in [0, 1]$. If these two conditions are satisfied, i.e., (1) $vol(\cm{S}^*) \leq \frac{1}{2}$ and (2) $\, \cm{S}^*$ is also the vertex set with optimal conductance among all the vertex sets containing $\cm{S}^*\setminus\cm{S}$, then with probability\footnote{the randomness comes from the sampling of $\cm{S}$} at least $\frac{1}{2}$, we have
\begin{equation}
    \Phi(pr(\alpha, \psi_{\cm{S}})) < O(\sqrt{\Phi(\cm{S}^*)})
\end{equation}
Specifically, when $vol(\cm{S}^*) \leq \frac{1}{3}$, then
\begin{equation}
    \Phi(pr(\alpha, \psi_{\cm{S}})) <  \sqrt{235\Phi(\cm{S}^{*})}
\end{equation}
(Proof in Appendix \ref{pf: conductance bound})
\end{theorem}

\begin{table}[t]
\setlength{\tabcolsep}{1mm}
\caption{Average Conductance ($\Phi$) and F1 on Local Clustering.}
\begin{center}
\scalebox{0.8}{
\begin{tabular}{l|c|cccc}
\toprule
Dataset & Metric & STAR++ & CLIQUE++ & DiffEq & HyperACL \\
\midrule
\multirow{2}{*}{DBLP-ML} 
  & $\Phi$ ($\downarrow$) & 0.2129 & \underline{0.2012} & 0.4705 & \textbf{0.1590} \\
  & F1 ($\uparrow$)    & 0.0918 & 0.0957             & \underline{0.0961} & \textbf{0.1125} \\
\midrule
\multirow{2}{*}{DBLP-CV} 
  & $\Phi$ ($\downarrow$) & 0.1704 & \underline{0.1576} & 0.3966 & \textbf{0.1104} \\
  & F1 ($\uparrow$)    & 0.2514 & \underline{0.2620} & 0.2307 & \textbf{0.2697} \\
\midrule
\multirow{2}{*}{DBLP-NLP} 
  & $\Phi$ ($\downarrow$) & 0.1045 & \underline{0.0949} & 0.3346 & \textbf{0.0559} \\
  & F1 ($\uparrow$)    & 0.3828 & \underline{0.4053} & 0.3259 & \textbf{0.4061} \\
\midrule
\multirow{2}{*}{DBLP-IR} 
  & $\Phi$ ($\downarrow$) & \underline{0.0737} & 0.0766 & 0.3260 & \textbf{0.0189} \\
  & F1 ($\uparrow$)    & \underline{0.4038} & 0.3997 & 0.2600 & \textbf{0.4802} \\
\bottomrule
\end{tabular}
}
\end{center}
\label{tb: average conductance and f1}
\end{table}

\section{Experiments}
\label{section: exp}

In this section, we demonstrate the effectiveness of our algorithm on real-world hypergraphs. We first describe the experimental settings and then discuss the experiment results. Details in Appendix \ref{ap: exp_details}.

\subsection{Experimental Setup}

\textbf{Datasets and Hypergraph Constructions.}
We construct 4 citation networks that are subgraphs of AMiner DBLP v14. For hypergraph construction, we use each vertex to represent a scholar and each hyperedge to represent a publication, connecting all its authors. Each hyperedge will be assigned a weight $w(e) = (\text{\# citations of paper }e) + 1$. Then, realistically, we assign more Edge-Dependent Vertex Weights for the leading authors and the last authors, and less EDVW for the middle authors. More details on the datasets and hypergraph construction are in Appendix \ref{ap: datasets}.

\textbf{Settings and Metrics.} For each dataset, we sample 50 seed vertex sets, which we call observations. For each observation, we first uniformly sample one organization within \{MIT, CMU, Stanford, UCB\}, then uniformly sample 5 authors (without duplication) affiliated with the sampled organization. Then, we run the methods to get local clusters and compare their quality in terms of (1) low conductance. A good local cluster should be a cohesive set of vertices; (2) a high F1 score with the set of all authors in the sampled organization. In real life, scholars usually coauthor more with others in the same academic institute. Hence, a good cluster should, to some extent, align with the institute label information \citep{DBLP:conf/www/LiuV0LG21}.

\textbf{Baselines.} We compare our HyperACL with the classic and strong STAR++ and CLIQUE++ expansion-based ACL local clustering methods~\citep{DBLP:conf/focs/AndersenCL06} (as expressed in Appendix~\ref{ap: baselines}), as well as DiffEq~\citep{DBLP:conf/kdd/Takai0IY20}, which is specifically tailored for non-EDVW hypergraph local clustering. It is proved in \citep{DBLP:conf/icml/ChitraR19} that, random walks on non-EDVW graphs are equivalent to some clique graph. 

\subsection{Experimental Results}

Figure \ref{fig: local results} shows the local clustering experiment results. in Table \ref{tb: average conductance and f1}, we report the average conductance and F1 score of the 50 observations on the local clustering task. \textbf{First}, in terms of both conductance and F1 score, HyperACL has the best performance on all the datasets. The purple lines, which represent our HyperACL, are significantly the lowest for conductance and generally the highest for F1. \textbf{Second}, CLIQUE++ and STAR++ can achieve sub-optimal results. \textbf{Third}, HyperACL is more stable because (1) for almost all the observations, our HyperACL can achieve a smaller conductance; (2) there are several observations in DBLP-IR where our HyperACL achieves a small conductance, but the baseline methods achieve relatively very large conductance.

We report the time per execution for the methods in the local clustering experiments in Figure \ref{fig: local times}. Generally, CLIQUE++ and HyperACL are better than STAR++ in terms of execution time. Compared to other methods, DiffEq is more stable. DiffEq achieves the best time performance on DBLP-ML, and achieves similar time performance on DBLP-CV and DBLP-NLP. However, DiffEq does not perform well on DBLP-NLP. Generally, our HyperACL has a similar time performance with CLIQUE++.

\subsection{Additional Experiments}
We put additional experiment results in the Appendix \ref{ap:additional_exp}, including (1) HyperACL on a larger dataset and beyond citation networks (\ref{ap:more_datasets}); (2) GeneralACL results on graph datasets (\ref{ap:graph_datasets}); (3) an ablation study on the early-stop mechanism (\ref{ap:early_stop}) with details of the time breakdown; (4) a hyperparameter study of early-stop steps (\ref{ap:early_stop_hyper}). (5) Weighting scheme ablation (\ref{ap:weighting_scheme}).

\section{Related Works}
\label{section: related works}
\noindent\textbf{Localized Graph Clustering.} Localized graph clustering \citep{DBLP:journals/csr/Schaeffer07} has been widely adopted across various data modalities \citep{DBLP:journals/pami/ShiM00}. Many existing methods rely on random walks \citep{DBLP:conf/focs/AndersenCL06, DBLP:journals/im/AndersenCL08, DBLP:conf/sigir/AvrachenkovDNPS08} or diffusion processes \citep{DBLP:journals/im/Chung09, DBLP:journals/ejc/ChungS18, DBLP:conf/icml/Fountoulakis0Y20} to identify communities based on localized connectivity patterns.
Recent developments have extended these methods to increasingly complex scenarios, including noisy data \citep{DBLP:conf/iclr/LucaF024}, attributed graphs \citep{DBLP:conf/icml/0002F23}, and dynamic graphs \citep{DBLP:conf/kdd/FuZH20}. We extend localized graph clustering to more general graph/hypergraph settings. 
\noindent\textbf{Hypergraphs.} A variety of techniques for hypergraph analysis have been proposed \citep{DBLP:journals/pami/GaoZLZDZ22, DBLP:journals/csur/AntelmiCPSSY24, lee2024survey}, though only a limited number address EDVW hypergraphs, whose spectral properties have been recently studied in \citep{li2024hypergraphs}. On the local clustering front, LQHD \citep{DBLP:conf/www/LiuV0LG21} introduced a strongly local diffusion algorithm for non-EDVW hypergraphs. Moreover, LQHD lacks any proof of optimality, limiting its theoretical robustness. 
DiffEq \citep{DBLP:conf/kdd/Takai0IY20} formulates hypergraph clustering using a differential equation framework and employs a different sweep cut approach compared to ours. In our experiments, we demonstrate that HyperACL outperforms DiffEq.
\noindent\textbf{PageRank.} PageRank \citep{page1999pagerank}, playing a crucial role in ranking systems, has been widely applied in graph analysis and mining. Some recent advances in this algorithm include theoretical optimality \citep{DBLP:conf/stoc/WangWW024}, more accelerated computation \citep{DBLP:conf/www/LiFH23, DBLP:journals/amc/ZhangTYZ23, DBLP:journals/tkde/YangWWWW24} and applications on diverse tasks \citep{ban2024pagerank, DBLP:conf/www/StoicaLC24}. Despite these advancements, most approaches focus on simple graph structures, overlooking more complex settings, as in this work.

\section{Conclusion}

In this work, we propose GeneralACL and HyperACL to handle weighted, directed, and self-looped graphs, as well as hypergraphs with edge-dependent vertex weights. Our algorithms provide theoretical guarantees, achieving quadratically optimal conductance. Experiments on diverse datasets confirm both the efficiency and effectiveness of our algorithms.


\section*{Acknowledgement}

This work is supported by National Science Foundation under Award No. IIS-2117902. The views and conclusions are those of the authors and should not be interpreted as representing the official policies of the funding agencies or the government.

\bibliographystyle{ACM-Reference-Format}
\bibliography{main.bib}

\clearpage
\appendix

\onecolumn
\textbf{\LARGE Appendix}

\addtocontents{toc}{\protect\setcounter{tocdepth}{2}}

\tableofcontents

\section{Useful Lemmas}

\begin{lemma}
    Denote $rpr(\alpha' , s)$ to be the solution of 
\begin{equation}
    rpr(\alpha', s) = \alpha' s + (1-\alpha') rpr(\alpha', s) P
\end{equation}
then $pr(\alpha, s) = rpr(\frac{2\alpha}{1+\alpha}, s)$. (Proof in Appendix \ref{pf: eols})
\label{lem: equivalent of lazy and standard}
\end{lemma}

\begin{lemma}
    $\forall \cm{S} \subseteq \cm{V}, p(S) \leq p[vol(S)]$.
\label{lem: ps and pvols}
(Proof in Appendix \ref{pf: ps and pvols})
\end{lemma}

\begin{lemma}
    Recall Definition \ref{df: |A|}, For any edge set $\cm{A} \in \cm{V} \times \cm{V}$, $p(\cm{A}) \leq p[|\cm{A}|]$.
\label{lem: pa and pa}
(Proof in Appendix \ref{pf: pa and pa})
\end{lemma}

\begin{lemma}
    Recall definition \ref{df: LSC} of $p[\cdot]$. For any starting distribution $s$ and any $k \in [0, 1]$,
\begin{equation}
    pr(\alpha, s)[k] \leq s[k]
\end{equation}
\label{lem: PageRank versus original}
(Proof in Appendix \ref{pf: PageRank versus original})
\end{lemma}

\section{Proofs}
\label{ap: Proof of Lemmas and Theorems in the Main Pages}

\subsection{Proof of Theorem \ref{thm: Phi(S) bound}}
\label{pf: Phi(S) bound}

\begin{proof}
Recall from the definition \ref{df: conductance}, \ref{df: volume of boundary} and \ref{df: volume of a set}, 

\begin{equation}
    \Phi_{\cm{H}}(\cm{S}) = \frac{|\partial\cm{S}|}{\min(vol(\cm{S}), vol(\bar{\cm{S}}))} = \frac{\sum_{u \in \cm{S}, v \in \cm{\bar{\cm{S}}}}\phi(u)P_{u,v}}{\min(\sum_{v \in \cm{S}}\phi(u), \sum_{u \in \bar{\cm{S}}}\phi(u))}
\end{equation}

From the definition of $P_{u, v}$ and $\phi(u)$, $\Phi(\cm{S})$ is non-negative. 

\begin{equation}
\begin{split}
    \sum_{u \in \cm{S}}\phi(u) &= \sum_{u \in \cm{S}, v \in \cm{V}} \phi(u)P_{u, v} \geq \sum_{u \in \cm{S}, v \in \bar{\cm{S}}} \phi(u)P_{u, v} \\
    \sum_{u \in \bar{\cm{S}}}\phi(u) &= \sum_{u \in \bar{\cm{S}}, v \in \cm{V}} \phi(u)P_{u, v} \geq \sum_{u \in \bar{\cm{S}}, v \in \cm{S}} \phi(u)P_{u, v} \overset{Equation \, \ref{thm: equality of in and out}}{=} \sum_{u \in \cm{S}, v \in \bar{\cm{S}}} \phi(u)P_{u, v} 
\end{split}
\end{equation}

Thus, $\min(\sum_{v \in \cm{S}}\phi(u), \sum_{u \in \bar{\cm{S}}}\phi(u)) \geq \sum_{u \in \cm{S}, v \in \cm{\bar{\cm{S}}}}\phi(u)P_{u,v}$ and $\Phi_{\cm{H}}(\cm{S}) \in [0, 1]$.
\end{proof}

\subsection{Proof of Lemma \ref{lem: equivalent of lazy and standard}}
\label{pf: eols}

\begin{proof}
Recall the definition of $M$ from \ref{df: pagerank}, and given that
\begin{equation}
    pr(\alpha, s) = \alpha s + (1-\alpha) pr(\alpha, S) M  
\end{equation}
it follows that
\begin{equation}
    pr(\alpha, s) = \alpha s + \frac{1-\alpha}{2} pr(\alpha, s) + \frac{1 - \alpha}{2}\cdot pr(\alpha, s)\cdot P
\end{equation}
and therefore,
\begin{equation}
    \frac{1+\alpha}{2}pr(\alpha, s) = \alpha s + \frac{1-\alpha}{2}pr(\alpha, s)\cdot P
\end{equation}
Then, we can give the final expression of $pr(\alpha,s)$ as follows:
\begin{equation}
    pr(\alpha,s) = \frac{2\alpha}{1+\alpha}\cdot s + \frac{1- \alpha}{1 + \alpha}\cdot pr(\alpha, s) \cdot P
\end{equation}
On the other hand,
\begin{equation}
    rpr \left(\frac{2\alpha}{1+\alpha}, S\right) = \frac{2\alpha}{1+\alpha} s + \frac{1-\alpha}{1+\alpha} pr(\alpha, s) P
\end{equation}
which are in the same function of $s, \alpha$ as $pr(\alpha,s)$.
So, because of the uniqueness, $pr(\alpha, S)$ is equal to $rpr \left(\frac{2\alpha}{1+\alpha}, S\right)$.
It is easy to verify that $\frac{2\alpha}{1+\alpha}$ is a bijective mapping for $\alpha \in [0,1]$.
Therefore, a lazy Random Walk is equivalent to standard PageRank up to a change of $\alpha$.

\end{proof}

\subsection{Proof of Lemma \ref{lem: ps and pvols}}
\label{pf: ps and pvols}

\begin{proof}
Recall from the definition \ref{df: LSC}, we can have this format of representation of L-S Curve:
\begin{equation}
    p[vol(\cm{S})] = \begin{cases}
        0 & vol(\cm{S}) = 0\\
        p(\cm{S}_j^p) & if \,\, \exists j \,\, s.t. \,\, vol(\cm{S})=vol(\cm{S}_j^p)\\
        p(\cm{S}_j^p) + (vol(\cm{S}) - vol(\cm{S}_j^p)) \frac{p(v_{j+1})}{\phi(v_{j+1})} & if \,\, \exists j \,\, s.t. \,\, vol(\cm{S}_j^p)<vol(\cm{S})<vol(\cm{S}_{j+1}^p)\\
        1 & vol(\cm{S}_{N_p}^p) < vol(\cm{S}) \leq 1 \\
    \end{cases}
\end{equation}

when $vol(\cm{S}) = 0$, the inequality holds as $p(\cm{S}) = p[vol(\cm{S})] = 0$;\\

when $vol(\cm{S})=vol(\cm{S}_j^p)$, then $p(\cm{S}) = p(\cm{S}_j^p)$, so the inequality holds as $p(\cm{S}) = p[vol(\cm{S})] = p(\cm{S}_j^p)$ for some $j$;\\

when $vol(\cm{S}_{N_p}^p) < vol(\cm{S}) \leq 1$, we can verify the inequality from the definition of $p$, as $p(\cm{S}) \leq 1$, so the inequality still holds.\\

For $vol(\cm{S}_j^p)<vol(\cm{S})<vol(\cm{S}_{j+1}^p)$, we can notice that $p[vol(\cm{S})]$ is a linear function of $vol(\cm{S})$ with slope $\frac{p(v_{j+1})}{\phi(v_{j+1})}$. Recall from the definition, we have proved that $p[k]$ is a continuous function. And 2 endpoints for the segment are $(vol(\cm{S}_j^p), p(\cm{S}_j^p))$ and $(vol(\cm{S}_{j+1}^p), p(\cm{S}_{j+1}^p))$, since $p(\cm{S})$ has no corresponding sweep set when $vol(\cm{S}_j^p)<vol(\cm{S})<vol(\cm{S}_{j+1}^p)$, so the inequality holds for this domain.

Overall, as $p(\cm{S})$ is not a continuous function, we can successfully prove that continuous $ p(S) \leq p[vol(S)]$.
\end{proof}

\subsection{Proof of Lemma \ref{lem: pa and pa}}
\label{pf: pa and pa}

\begin{proof}
Given \( B_j = \{ (u,v) \in E | u \in \cm{S}_j^P \} \), and \( f(u,v) =  \phi(u) \cdot p(u, v)\),  with $g(u) = \frac{p(u)}{\phi(u)}$ if \( (u,v) \in B_j \), we are going to prove from contradiction:

If
\begin{equation}
\label{contradicton assumption}
    \sum_{(u,v) \in A} g(u) \cdot f(u,v) > \sum_{(u,v) \in B_j} g(u) \cdot f(u,v)
\end{equation}

Then, we have 
\begin{equation}
    \frac{\sum_{(u,v) \in A} g(u) \cdot f(u,v)}{\sum_{(u,v) \in A} f(u,v)} > \frac{\sum_{(u,v) \in B_j} g(u) \cdot f(u,v)}{\sum_{(u,v) \in B_j} f(u,v)} \geq g(v_j)
\end{equation}

Then, there exists $(u_1,v_1) \in A$ with $u_1 \in S_{j-1}$ such that letting $A^1 = A \setminus \{(u_1,v_1)\}$, $B^1_j = B_j \setminus \{(u_1,v_1)\}$, we have:

\begin{equation}
    \sum_{(u,v) \in A^1} f(u,v) = \sum_{(u,v) \in B^1_j} f(u,v)
\end{equation}

\begin{equation}
    \sum_{(u,v) \in A^1} g(u) \cdot f(u,v) > \sum_{(u,v) \in B^1_j} g(u) \cdot f(u,v)
\end{equation}
\begin{equation}
    \frac{\sum_{(u,v) \in A^1} g(u) \cdot f(u,v)}{\sum_{(u,v) \in A^1} f(u,v)} > \frac{\sum_{(u,v) \in B^1_j} g(u) \cdot f(u,v)}{\sum_{(u,v) \in B^1_j} f(u,v)} \geq g(v_j)
\end{equation}

Therefore, similarly, for some edge $(u_2,w_2) \in A^1$, where $u \in S_{j-1}^{p}$, letting $A^2 = A^1 \setminus \{(u,w)\}$; $B^2_j = B^1_j \setminus \{(u,w)\}$, It holds that
\begin{equation}
    \sum_{(u,v) \in A^2} f(u,v) = \sum_{(u,v) \in B^2_j} f(u,v)
\end{equation}

and
\begin{equation}
    \sum_{(u,v) \in A^2} g(u) \cdot f(u,v) > \sum_{(u,v) \in B^2_j} g(u) \cdot f(u,v)
\end{equation}

Consequently, we keep this iteration until for some $k$, $A^k = \phi$ or $B_j^k = \phi$, as for empty set, and the definition that $f(u, v) \geq 0$
\begin{equation}
    \sum_{(u,v) \in A^k} f(u,v) = \sum_{(u,v) \in B^k_j} f(u,v) = 0
\end{equation}
Then, we can get
\begin{equation}
    \sum_{(u,v) \in A^1} k \cdot g(u) \cdot f(u,v) = \sum_{(u,v) \in B^1_j} k \cdot g(u) \cdot f(u,v) = 0
\end{equation}
which implies
\begin{equation}
    \sum_{(u,v) \in A} g(u) \cdot f(u,v) = \sum_{(u,v) \in B_j} g(u) \cdot f(u,v)
\end{equation}
which is contradictory to the assumption (Equation \ref{contradicton assumption}).
\end{proof}

\subsection{Proof of Lemma \ref{lem: pms in out}}
\label{pf: pms in out}

\begin{proof}
    For any node $u \in \cm{V}$, we have $\phi(u) = \sum_{v\in \cm{V}}\phi(v)P_{v, u}$

\begin{equation}
    (pM)(\{u\}) = (\frac{1}{2}pI + \frac{1}{2}pP)(\{u\}) = \frac{1}{2}p(u) + \frac{1}{2}pP(u) = \frac{1}{2}p(u) + \frac{1}{2} \sum_{v \in \cm{V}} p(v) P_{v, u}
\end{equation}

\begin{equation}
    p(out(\{u\})) = \sum_{v \in \cm{V}}\phi(u)p(u, v) = \sum_{v \in \cm{V}}\phi(u)\frac{p(u)}{\phi(u)}P_{u, v} = \sum_{v \in \cm{V}}p(u)P_{u, v} = p(u)
\end{equation}

\begin{equation}
    p(in(\{u\})) = \sum_{v \in \cm{V}}\phi(v)p(v, u) = \sum_{v \in \cm{V}}\phi(v)\frac{p(v)}{\phi(v)}P_{v, u} = \sum_{v \in \cm{V}}p(v)P_{v, u} = pP(u)
\end{equation}

Therefore, $(pM)(\{u\}) = \frac{1}{2} p(in(\{u\})) + \frac{1}{2} p(out(\{u\}))$ and

\begin{equation}
\begin{split}
        (pM)(\cm{S}) &= \sum_{u \in \cm{S}}(pM)(\{u\}) = \sum_{u \in \cm{S}}\frac{1}{2} p(in(\{u\})) + \frac{1}{2} p(out(\{u\}))\\
        &= \frac{1}{2} \sum_{u \in \cm{S}}p(in(\{u\})) + \frac{1}{2} \sum_{u \in \cm{S}}p(out(\{u\}))\\
        &= \frac{1}{2} p(in(\cm{S})) + \frac{1}{2}p(out(\cm{S}))\\
        &= \frac{1}{2}p(in(\cm{S}) \cup out(\cm{S})) + \frac{1}{2}p(in(\cm{S})\cap p(out(\cm{S}))\\
\end{split}
\end{equation}

The last equation holds because for any two edge sets $\cm{A}$, $\cm{B}$, let $k(u, v) = p(u)P_{u, v}$, then,
\begin{equation}
    p(\cm{A}) + p(\cm{B})  = \sum_{(u, v) \in \cm{A}}k(u, v) + \sum_{(u, v) \in \cm{B}}k(u, v) = \sum_{(u, v) \in \cm{A}\cup\cm{B}}k(u, v) + \sum_{(u, v) \in \cm{A}\cap\cm{B}}k(u, v) = p(\cm{A}\cup\cm{B}) + p(\cm{A}\cap\cm{B})
\end{equation}

\end{proof}

\subsection{Proof of Lemma \ref{lem: ababab}}
\label{pf: ababab}

\begin{proof}
\begin{equation}
\begin{split}
    |\cm{A}| + |\cm{B}|  &= \sum_{(u, v) \in \cm{A}}\phi(u)P_{u, v} + \sum_{(u, v) \in \cm{B}}\phi(u)P_{u, v} \\
    &= \sum_{(u, v) \in \cm{A}\cup\cm{B}}\phi(u)P_{u, v} + \sum_{(u, v) \in \cm{A}\cap\cm{B}}\phi(u)P_{u, v} \\
    &= |\cm{A}\cup\cm{B}| + |\cm{A}\cap\cm{B}|
\end{split}
\end{equation}
\end{proof}

\subsection{Proof of Lemma \ref{lem: in out vol}}
\label{pf: in out vol}

\begin{proof}
\begin{equation}
\begin{split}
    |in(\cm{S})| + |out(\cm{B})|  &= |\{(u, v)| u\in \cm{V}, v\in \cm{S}\}| + |\{(u, v)| u\in \cm{S}, v\in \cm{V}\}| \\
    &= \sum_{v\in {\cm{V}}, u \in \cm{S}} \phi(v)P_{v, u} + \sum_{u \in \cm{S}, v\in {\cm{V}}} \phi(u)P_{u, v} \\
    &=  \sum_{u \in \cm{S}} \sum_{v\in {\cm{V}}} \phi(v)P_{v, u} + \sum_{u \in \cm{S}} \sum_{v\in {\cm{V}}} \phi(u)P_{u, v} \\
    &= \sum_{u \in \cm{S}} \phi(u) + \sum_{u \in \cm{S}} \phi(u) \sum_{v\in {\cm{V}}} P_{u, v}\\
    &= \sum_{u \in \cm{S}} \phi(u) + \sum_{u \in \cm{S}} \phi(u) \sum_{v\in {\cm{V}}} P_{u, v}\\
    &= \sum_{u \in \cm{S}} \phi(u) + \sum_{u \in \cm{S}} \phi(u) \\
    &= vol(S) + vol(S)\\
    &= 2vol(S)
\end{split}
\end{equation}

\end{proof}

In fact, from the above proof, we have a stronger conclusion that $|in(\cm{S})| = |out(\cm{S})| =  vol(\cm{S})$.

\subsection{Proof of Lemma \ref{lem: in out partial}}
\label{pf: in out partial}

\begin{proof}
\begin{equation}
\begin{split}
    |in(\cm{S})\cup out(\cm{B})| &+ |in(\cm{B})\cap out(\cm{B})| = \sum_{(u\in {\cm{S}}, v \in \cm{V}) \lor (u\in {\cm{V}}, v \in \cm{S})} \phi(u)P_{u, v} - \sum_{(u \in \cm{S}, v\in \cm{V}) \land (u \in \cm{V}, v\in \cm{S})} \phi(u)P_{u, v} \\
    &= (\sum_{u\in \bar{\cm{S}}, v \in \cm{S}} \phi(u)P_{u, v} + \sum_{u\in {\cm{S}}, v \in \bar{\cm{S}}} \phi(u)P_{u, v} + \sum_{u\in \cm{S}, v \in \cm{S}} \phi(u)P_{u, v}) - \sum_{u \in \cm{S}, v\in {\cm{S}}} \phi(u)P_{u, v} \\
    &= \sum_{u\in \bar{\cm{S}}, v \in \cm{S}} \phi(u)P_{u, v} + \sum_{u\in {\cm{S}}, v \in \bar{\cm{S}}} \phi(u)P_{u, v}\\
    &= |\partial \cm{S}| + |\partial \cm{S}| \\
    &= 2 |\partial \cm{S}|
\end{split}
\end{equation}
\end{proof}

\subsection{Proof of Lemma \ref{lem: in out vol and partial}}
\label{pf: in out vol and partial}

\begin{proof}
    From Lemma \ref{lem: ababab} and Lemma \ref{lem: in out vol},
\begin{equation}
    |in(\cm{S})\cup out(\cm{S})| + |in(\cm{S})\cap out(\cm{S})| = |in(\cm{S})| + |out(\cm{S})| = 2vol(\cm{S})
\end{equation}

    From Lemma \ref{lem: in out partial},
\begin{equation}
    |in(\cm{S})\cup out(\cm{S})| - |in(\cm{S})\cap out(\cm{S})| = 2|\partial \cm{S}|
\end{equation}

The above two equations give
\begin{equation}
\begin{split}
    |in(\cm{S})\cup out(\cm{S})| &= vol(\cm{S}) + |\partial \cm{S}| \\
    |in(\cm{S})\cap out(\cm{S})| &= vol(\cm{S}) - |\partial \cm{S}| \\
\end{split}
\end{equation}

\end{proof}

\subsection{Proof of Lemma \ref{lem: pvol pvol+-partial}}
\label{pf: pvol pvol+-partial}

\begin{proof}
$\forall \cm{S} \in \cm{V}$,
\begin{equation}
\begin{split}
    p(\cm{S}) &= pr(\alpha, s) =  (\alpha s + (1-\alpha pM))(\cm{S})\\
         &=  \alpha s(\cm{S}) + ((1-\alpha) pM) (\cm{S}) \\
         &\overset{Lemma \, \ref{lem: pms in out}}{=} \alpha s(\cm{S}) + (1-\alpha) \frac{1}{2}(p(in(\cm{S}) \cup out(\cm{S})) + p(in(\cm{S})\cap out(\cm{S}))) 
\end{split}
\end{equation}

Recall from definition \ref{df: LSC} that $p[vol(\cm{S}_j^p)] = p(\cm{S}_j^p)$,

\begin{equation}
\begin{split}
    p[vol(\cm{S}_j^p)] &= p(\cm{S}_j^p) \\
         &=  \alpha s(\cm{S}_j^p) + (1-\alpha) \frac{1}{2}(p(in(\cm{S}_j^p) \cup out(\cm{S}_j^p)) + p(in(\cm{S}_j^p)\cap out(\cm{S}_j^p)))  \\
         &\overset{Lemma \, \ref{lem: pa and pa}}{\leq} \alpha s(\cm{S}_j^p) + (1-\alpha) \frac{1}{2}(p[|in(\cm{S}_j^p) \cup out(\cm{S}_j^p)|] + p[|in(\cm{S}_j^p)\cap out(\cm{S}_j^p)|]) \\
         &\overset{Lemma \, \ref{lem: in out vol and partial}}{=} \alpha s(\cm{S}_j^p) + (1-\alpha) \frac{1}{2}(p[vol(\cm{S}_j^p) + |\partial S_j^p|] + p[vol(\cm{S}_j^p) - |\partial S_j^p|]) \\
         &\overset{Lemma \, \ref{lem: ps and pvols}}{\leq} \alpha s[vol(\cm{S}_j^p)] + (1-\alpha) \frac{1}{2}(p[vol(\cm{S}_j^p) + |\partial S_j^p|] + p[vol(\cm{S}_j^p) - |\partial S_j^p|]) \\
\end{split}
\end{equation}

\end{proof}

\subsection{Proof of Theorem \ref{thm: either or}}
\label{pf: either or}

\begin{proof}
    Let $k_j = vol(\cm{S}_j^p), \bar{k_j} = \min (k_j, 1-k_j)$. Let
\begin{equation}
    f_t(k) = \gamma + \alpha t + \sqrt{\frac{1}{\theta}}\sqrt{\min(k, 1-k)} (1 - \frac{\sigma^2}{8})^t
\end{equation}

    For $k=0$ or $k=1$, $p[k] - k = 0 \leq f_t(k)$, and the ``either" side holds.

    If there is no sweep cut satisfying the two properties at the ``or" side, we aim to prove the ``either" side holds. In other words, we aim to prove that, for all $t \geq 0$ and $k \in [\theta, 1-\theta]$,

\begin{equation}
    p[k] - k \leq f_t(k) 
\label{eq: inequality 1 to prove in theorem either or}
\end{equation}

We prove this by induction. When $t = 0$, $f_t(k) = \gamma + \sqrt{\frac{1}{\theta}}\sqrt{\min(k, 1-k)} \geq \sqrt{\frac{1}{\theta}}\sqrt{\min(k, 1-k)}$,

\begin{equation}
    p[k] - k \leq p[k] \leq 1 = \sqrt{\frac{1}{\theta}}\sqrt{\theta} \overset{k\in [\theta, 1-\theta]}{\leq}  \sqrt{\frac{1}{\theta}}\sqrt{\min(k, 1-k)}
\end{equation}

\textbf{Assume for the induction that}, for $t$, the inequality \ref{eq: inequality 1 to prove in theorem either or}: $p[k] - k \leq f_t(k)$ holds. we aim to prove that the inequality \ref{eq: inequality 1 to prove in theorem either or} also holds for $t+1$, i.e., $p[k] - k \leq f_{t+1}(k)$. 

Note the fact that the sum of two concave functions is still concave. With respect to $k$, since $\sqrt{\min (k, 1-k)}$ is concave, we have $f_{t+1}(k)$ is concave, and furthermore $f_{t+1}(k) + k$ is concave. To show $p[k] - k \leq f_{t+1}(k)$, it suffices to prove 
\begin{equation}
    p[k_j] \leq f_{t+1}(k_j) + k_j
\label{eq: inequality 2 to prove in theorem either or}
\end{equation}

Then, since $f_{t+1}(k) + k$ is concave and $p[k]$ is piecewise linear, $\forall k_j < k < k_{j+1}$ we have $p[k] \leq f_{t+1}(k) + k$.

Therefore, to finish the induction, we only need to prove the inequality \ref{eq: inequality 2 to prove in theorem either or}. Since there is no sweep cut satisfying the two properties, $\forall j\in [1, |Supp(p)|]$ (recall definition \ref{df: sweep}), $\cm{S}_j^p$ does not satisfy Property 1 or Property 2. 

\textbf{If} $\cm{S}_j^p$ does not satisfy property 2, then $\forall t \in \mathbb{N}_+$, 
\begin{equation}
\begin{split}
    & \qquad \quad p(\cm{S}_j^p) - vol(\cm{S}_j^p) \leq \gamma + \alpha t + \sqrt{\frac{1}{\theta}}\sqrt{\min(vol(\cm{S}_j^p), 1-vol(\cm{S}_j^p))} (1 - \frac{\sigma^2}{8})^t \\
    & \iff p[vol(\cm{S}_j^p)] - vol(\cm{S}_j^p) \leq f_t(vol(\cm{S}_j^p))\\
    & \iff p[k_j] - k_j \leq f_t(k_j)
\end{split}
\label{eq: either or condition 1}
\end{equation}

\textbf{Else}, $\cm{S}_j^p$ does not satisfy property 1. Then, $\Phi(\cm{S}_j^p) \geq \sigma$. Since $p[k]$ is a monotonically non-decreasing piecewise linear concave function,
\begin{equation}
    \mathrm{when}\,\, b \geq c,~  p[a-b] + p[a+b] \leq p[a-c] + p[a+c]
\label{eq: concavity of p[k]}
\end{equation}

\begin{equation}
\begin{split}
\begin{array}{c c l}
    p[k_j]  & \overset{k_j = vol(\cm{S}_j^p)}{=} & p[vol(\cm{S}_j^p)] \\
            &\overset{Lemma \, \ref{lem: pvol pvol+-partial}}{\leq}& \alpha s[vol(\cm{S}_j^p)] + (1-\alpha) \frac{1}{2}(p[vol(\cm{S}_j^p) + |\partial \cm{S}_j^p|] + p[vol(\cm{S}_j^p) - |\partial \cm{S}_j^p|]) \\
            &\overset{switch \,\, terms}{=}& \alpha s[vol(\cm{S}_j^p)] + (1-\alpha) \frac{1}{2}(p[vol(\cm{S}_j^p) - |\partial \cm{S}_j^p|] + p[vol(\cm{S}_j^p) + |\partial \cm{S}_j^p|]) \\
            &\overset{Definition \, \ref{df: volume of boundary}}{=}& \alpha s[vol(\cm{S}_j^p)] + (1-\alpha) \frac{1}{2}(p[vol(\cm{S}_j^p) - \Phi(\cm{S}_j^p)\bar{k_j}] + p[vol(\cm{S}_j^p) + \Phi(\cm{S}_j^p)\bar{k_j}])\\
            &\overset{Equation \, \ref{eq: concavity of p[k]}}{\leq}& \alpha s[vol(\cm{S}_j^p)] + (1-\alpha) \frac{1}{2}(p[vol(\cm{S}_j^p) - \sigma\bar{k_j}] + p[vol(\cm{S}_j^p) + \sigma\bar{k_j}])\\
            &\overset{\alpha \in [0, 1], s[k] \leq 1}{\leq}& \alpha + \frac{1}{2}(p[vol(\cm{S}_j^p) - \sigma\bar{k_j}] + p[vol(\cm{S}_j^p) + \sigma\bar{k_j}])\\  
            & \overset{k_j = vol(\cm{S}_j^p)}{=} & \alpha + \frac{1}{2}(p[k_j - \sigma\bar{k_j}] + p[k_j + \sigma\bar{k_j}])\\  
            &\overset{induction \,\,hypothesis}{\leq}& \alpha + \frac{1}{2}(f_t(k_j - \sigma\bar{k_j}) + (k_j - \sigma\bar{k_j}) + f_t(k_j + \sigma\bar{k_j}) + (k_j + \sigma\bar{k_j}))\\    
            & = & \alpha + k_j + \frac{1}{2}(f_t(k_j - \sigma\bar{k_j}) + f_t(k_j + \sigma\bar{k_j}))\\
\end{array}
\end{split}
\end{equation}

Therefore,
\begin{equation}
\begin{split}
    p[k_j] - k_j &\leq \alpha + \frac{1}{2}(f_t(k_j - \sigma\bar{k_j}) + f_t(k_j + \sigma\bar{k_j})) \\
    &= \alpha + \gamma + \alpha t + \frac{1}{2}\sqrt{\frac{1}{\theta}}(1-\frac{\sigma^2}{8})^t(\sqrt{\min(k_j - \sigma\bar{k_j}, 1- k_j + \sigma\bar{k_j})} + \sqrt{\min(k_j + \sigma\bar{k_j}, 1 - k_j - \sigma\bar{k_j})})\\
\end{split}
\label{eq: equation to continue in theorem either or}
\end{equation}

$\mathbf{1^\circ \,\,}$ When $k_j \in[0, \frac{1}{2}], \bar{k_j} = k_j$. 
\begin{equation}
\begin{split}
    &\quad \,\, \sqrt{\min(k_j - \sigma\bar{k_j}, 1- k_j + \sigma\bar{k_j})} + \sqrt{\min(k_j + \sigma\bar{k_j}, 1 - k_j - \sigma\bar{k_j})}\\
    &= \sqrt{\min(\bar{k_j} - \sigma\bar{k_j}, 1- k_j + \sigma\bar{k_j})} + \sqrt{\min(\bar{k_j} + \sigma\bar{k_j}, 1 - k_j - \sigma\bar{k_j})}\\
    &\leq \sqrt{\bar{k_j} - \sigma \bar{k_j}} + \sqrt{\bar{k_j} + \sigma \bar{k_j}}\\
\end{split}
\end{equation}

$\mathbf{2^\circ \,\,}$ When $k_j \in[\frac{1}{2}, 1], \bar{k_j} = 1 - k_j.$
\begin{equation}
\begin{split}
    &\quad \,\, \sqrt{\min(k_j - \sigma\bar{k_j}, 1- k_j + \sigma\bar{k_j})} + \sqrt{\min(k_j + \sigma\bar{k_j}, 1 - k_j - \sigma\bar{k_j})}\\
    &= \sqrt{\min(1 - \bar{k_j} - \sigma \bar{k_j}, \bar{k_j} + \sigma \bar{k_j})} + \sqrt{\min(1 - \bar{k_j} + \sigma \bar{k_j}, \bar{k_j} - \sigma \bar{k_j})}\\
    &\leq \sqrt{\bar{k_j} + \sigma \bar{k_j}} + \sqrt{\bar{k_j} - \sigma \bar{k_j}}\\
    &= \sqrt{\bar{k_j} - \sigma \bar{k_j}} + \sqrt{\bar{k_j} + \sigma \bar{k_j}}\\
\end{split}
\end{equation}

Therefore, $\forall k_j \in [0, 1], \sqrt{\min(k_j - \sigma\bar{k_j}, 1- k_j + \sigma\bar{k_j})} + \sqrt{\min(k_j + \sigma\bar{k_j}, 1 - k_j - \sigma\bar{k_j})} \leq \sqrt{\bar{k_j} - \sigma \bar{k_j}} + \sqrt{\bar{k_j} + \sigma \bar{k_j}}$. Continue with Equation \ref{eq: equation to continue in theorem either or}, 

\begin{equation}
\begin{split}
    p[k_j] - k_j &\leq \alpha + \gamma + \alpha t + \frac{1}{2}(1-\frac{\sigma^2}{8})^t\sqrt{\frac{1}{\theta}}(\sqrt{\min(k_j - \sigma\bar{k_j}, 1- k_j + \sigma\bar{k_j})} + \sqrt{\min(k_j + \sigma\bar{k_j}, 1 - k_j - \sigma\bar{k_j})})\\
    &\leq \alpha + \gamma + \alpha t + \frac{1}{2}(1-\frac{\sigma^2}{8})^t\sqrt{\frac{1}{\theta}}(\sqrt{\bar{k_j} - \sigma \bar{k_j}} + \sqrt{\bar{k_j} + \sigma \bar{k_j}})\\
\label{eq: equation to continue 2 in theorem either or}
\end{split}
\end{equation}

Consider the Taylor series 
\begin{equation}
\begin{split}
    \sqrt{1+x} &= (1+x)^\frac{1}{2} = \sum_{n=0}^\infty \binom{\frac{1}{2}}{n} \cdot x^n \\
    \sqrt{1-x} &= (1+(-x))^\frac{1}{2} = \sum_{n=0}^\infty \binom{\frac{1}{2}}{n} \cdot (-x)^n \\
\end{split}
\end{equation}

\begin{equation}
\begin{split}
    \sqrt{1-x} + \sqrt{1+x} &= \sum_{n=0}^\infty \binom{\frac{1}{2}}{n} \cdot (-x)^n + \sum_{n=0}^\infty \binom{\frac{1}{2}}{n} \cdot x^n = 2\sum_{m=0}^\infty \binom{\frac{1}{2}}{2m} \cdot x^{2m} \\
     &\overset{\forall m\geq 2, \binom{\frac{1}{2}}{2m}\leq 0, x^{2m}\geq 0}{\leq}     2(\binom{\frac{1}{2}}{0} \cdot x^0 + \binom{\frac{1}{2}}{2} \cdot x^2)\\
     &= 2(1 + \frac{\frac{1}{2}\cdot (-\frac{1}{2})}{2} \cdot x^2) \\
     &= 2(1 - \frac{x^2}{8})
\end{split}
\end{equation}

Continue with Equation \ref{eq: equation to continue 2 in theorem either or} 
\begin{equation}
\begin{split}
    p[k_j] - k_j &\leq \alpha + \gamma + \alpha t + \frac{1}{2}(1-\frac{\sigma^2}{8})^t\sqrt{\frac{1}{\theta}}(\sqrt{\bar{k_j} - \sigma \bar{k_j}} + \sqrt{\bar{k_j} + \sigma \bar{k_j}})\\
    &= \gamma + \alpha (t+1) + \frac{1}{2}(1-\frac{\sigma^2}{8})^t\sqrt{\frac{1}{\theta}}\sqrt{\bar{k_j}}(\sqrt{1 - \sigma} + \sqrt{1 + \sigma)} \\
    & \overset{Equation \,\, \ref{eq: equation to continue 2 in theorem either or}}{\leq}   \gamma + \alpha (t+1) + \frac{1}{2}(1-\frac{\sigma^2}{8})^t\sqrt{\frac{1}{\theta}}\sqrt{\bar{k_j}}\cdot 2(1 - \frac{\sigma^2}{8})\\
    &= \gamma + \alpha (t+1) + (1-\frac{\sigma^2}{8})^{t+1}\sqrt{\frac{1}{\theta}}\sqrt{\bar{k_j}}\cdot \\
    &= \gamma + \alpha (t+1) + \sqrt{\frac{1}{\theta}}\sqrt{\min(k_j, 1-k_j)}(1-\frac{\sigma^2}{8})^{t+1}\\
    &= f_{t+1}(k_j)
\label{eq: either or condition 2}
\end{split}
\end{equation}

Combining equations \ref{eq: either or condition 1} and \ref{eq: either or condition 2}, no matter which property the sweep cut does not satisfy, we always have $p[k_j] - k_j \leq f_{t+1}(k_j)$, which completes the induction step.
\end{proof}

Though Theorem \ref{thm: either or} describes an ``either or" relation, we can obtain an upper bound by setting $\sigma = \Phi(pr(\alpha, s)) = \mathop{\min}_{j \in [1, N_p]} \Phi (\cm{S}_j^{pr(\alpha, s)}) = \mathop{\min}_{j \in [1, N_p]} \Phi (\cm{S}_j^p)$. Then, the "or" case cannot hold. Therefore the "either" case holds and we obtain an upper bound of the disagreement $p[k] - k$. Moreover, this upper bound is associated with $\sigma = \Phi(pr(\alpha, s))$.

\subsection{Proof of Theorem \ref{thm: conductance_satisfying_shrunk}}
\label{pf: conductance_satisfying_shrunk}
\begin{proof}
    Recall from Theorem \ref{thm: Phi(S) bound} and Definition \ref{df: optimal conductance of a distribution}, we have $\Phi(pr(\alpha, s)) \in [0, 1].$ In Theorem \ref{thm: either or}, let $\sigma = \Phi(pr(\alpha, s)), \gamma = 0, p = pr(\alpha, s)$ is a lazy PageRank vector. Then, for the constant $\theta = \min(vol(\cm{S}), 1 - vol(\cm{S})) \in [0, \frac{1}{2}]$,
\textbf{either} the following bound holds for any integer $t$ and any $k \in [\theta, 1-\theta] \cup \{0, 1\}$
\begin{equation}
    p[k] - k \leq \alpha t + \sqrt{\frac{1}{\theta}}\sqrt{\min(k, 1-k)} (1 - \frac{\sigma^2}{8})^t
\label{eq: using theorem either or in the shrunk version}
\end{equation}
    \textbf{or} there exist a sweep cut $\cm{S}_j^p, j \in [1, N_p]$, with the following properties,
\begin{equation}
\begin{split}
    & \mathrm{Property}\, 1.\, \Phi(\cm{S}_j^p) < \sigma\\
    & \mathrm{Property}\, 2.\, \exists t \in \mathbb{N}_+, p(\cm{S}_j^p) - vol(\cm{S}_j^p) > \alpha t + \sqrt{\frac{1}{\theta}}\sqrt{\min(vol(\cm{S}_j^p), 1-vol(\cm{S}_j^p))} (1 - \frac{\sigma^2}{8})^t
\end{split}
\end{equation}

As we set $\sigma = \Phi(pr(\alpha, s)) = \mathop{\min}_{j \in [1, N_p]} \Phi (\cm{S}_j^{pr(\alpha, s)}) = \mathop{\min}_{j \in [1, N_p]} \Phi (\cm{S}_j^p)$, the ``or" case cannot hold. Therefore, the ``either" case holds and  $\forall t \in \mathbb{N}_+$,
\begin{equation}
\begin{split}
    \delta &= pr(\alpha, s)(\cm{S}) - vol(\cm{S}) \overset{Lemma \, \ref{lem: ps and pvols}}{\leq} pr(\alpha, s)[vol(\cm{S})] - vol(\cm{S})\\
    &\overset{Equation \, \ref{eq: using theorem either or in the shrunk version}}{\leq} \alpha t + \sqrt{\frac{1}{\theta}}\sqrt{\min(vol(\cm{S}), 1-vol(\cm{S}))} (1 - \frac{\sigma^2}{8})^t \\
    & = \alpha t + \sqrt{\frac{1}{\min(vol(\cm{S}), 1 - vol(\cm{S}))}}\sqrt{\min(vol(\cm{S}), 1 - vol(\cm{S}))} (1 - \frac{\sigma^2}{8})^t   \\
    &= \alpha t + (1 - \frac{\sigma^2}{8})^t
\label{eq: shrunk to continue 1}
\end{split}
\end{equation}

$\forall z \leq \frac{1}{e}, \ln\frac{1}{z} \geq 1$, let $t_z = \frac{8}{\sigma^2}\ln\frac{1}{z}, t = \lceil t_z \rceil \geq t_z$, then,

\begin{equation}
\begin{split}
    \delta
    &\overset{Equation \, \ref{eq: shrunk to continue 1}}{\leq} \alpha t + (1 - \frac{\sigma^2}{8})^t \\
    &= \alpha \lceil \frac{8}{\sigma^2}\ln\frac{1}{z} \rceil + (1 - \frac{\sigma^2}{8})^{\lceil \frac{8}{\sigma^2}\ln\frac{1}{z} \rceil}\\
    &\leq  \alpha \lceil \frac{8}{\sigma^2}\ln\frac{1}{z} \rceil + (1 - \frac{\sigma^2}{8})^{ \frac{8}{\sigma^2}\ln\frac{1}{z}}\\
    &\leq  \alpha \lceil \frac{8}{\sigma^2}\ln\frac{1}{z} \rceil + (\frac{1}{e})^{ \ln\frac{1}{z}}\\
    &\leq  \alpha \lceil \frac{8}{\sigma^2}\ln\frac{1}{z} \rceil + z\\
    &=  \alpha ( \frac{8}{\sigma^2}\ln\frac{1}{z} + 1) + z\\
    &<  \alpha ( \frac{9}{\sigma^2}\ln\frac{1}{z}) + z\\
\end{split}
\end{equation}

Therefore, if $\delta > z$,
\begin{equation}
    \sigma \leq \sqrt{\frac{9\alpha \ln\frac{1}{z}}{\delta - z}}
\end{equation}

\end{proof}

\subsection{Proof of Lemma \ref{lem: PageRank versus original}}
\label{pf: PageRank versus original}
\begin{proof}
    Let $p = pr(\alpha, s)$. From Lemma \ref{lem: pvol pvol+-partial}, $\forall j\in [1, N_p]$, 
\begin{equation}
\begin{split}
    p[vol&(\cm{S}_j^p)] \leq \alpha s[vol(\cm{S}_j^p)] + (1-\alpha) \frac{1}{2}(p[vol(\cm{S}_j^p) + |\partial \cm{S}_j^p|] + p[vol(\cm{S}_j^p) - |\partial \cm{S}_j^p|]) \\
    & \overset{Concavity \, of \, p[k]}{\leq} \alpha s[vol(\cm{S}_j^p)] + (1-\alpha) \frac{1}{2}(p[vol(\cm{S}_j^p) + 0] + p[vol(\cm{S}_j^p) - 0]) \\
    &= \alpha s[vol(\cm{S}_j^p)] + (1-\alpha) (p[vol(\cm{S}_j^p)]) \\
    \iff & p[vol(\cm{S}_j^p)] - (1-\alpha) (p[vol(\cm{S}_j^p)]) \leq \alpha s[vol(\cm{S}_j^p)]\\
    \iff & p[vol(\cm{S}_j^p)] \leq s[vol(\cm{S}_j^p)]\\
\end{split}
\end{equation}
Since $s[k]$ is concave, $p[k]$ is piecewise linear, and the breakpoints $(vol(\cm{S}_j^p), p[vol(\cm{S}_j^p)])$ are bounded by $(vol(\cm{S}_j^p), s[vol(\cm{S}_j^p)])$, i.e., for each $k_j = vol(\cm{S}_j^p), p[k_j] \leq s[k_j]$, we have 

\begin{equation}
    p[k] \leq s[k] \iff pr(\alpha, s) \leq s[k] \,\, \forall k \in [0, 1].
\end{equation}

\end{proof}

\subsection{Proof of Lemma \ref{lem: partial_in = partial_out = partialS}}
\label{pf: partial_in = partial_out = partialS}

\begin{proof}
    Recall from the proof of Lemma \ref{lem: in out vol}, we have 
\begin{equation}
\begin{split}
    & \sum_{u \in \cm{S}, v \in \cm{V}}\phi(u) P_{u, v} = |out(\cm{S})| = vol(\cm{S})\\
    & \sum_{u \in \cm{V}, v \in \cm{S}}\phi(u) P_{u, v} = \sum_{u \in \cm{S}, v \in \cm{V}}\phi(v) P_{v, u} = |in(\cm{S})| = vol(\cm{S})\\
\end{split}
\end{equation}

\begin{equation}
\begin{split}
    & |out(S)| - \sum_{u \in \cm{S}, v \in \cm{S}}\phi(u) P_{u, v} = \sum_{u \in \cm{S}, v \in \cm{V}}\phi(u) P_{u, v} - \sum_{u \in \cm{S}, v \in \cm{S}}\phi(u) P_{u, v} = \sum_{u \in \cm{S}, v \in \bar{\cm{S}}}\phi(u) P_{u, v} = |\partial_{out}(\cm{S})| \\
    & |in(S)| - \sum_{u \in \cm{S}, v \in \cm{S}}\phi(u) P_{u, v} = \sum_{u \in \cm{V}, v \in \cm{S}}\phi(u) P_{u, v} - \sum_{u \in \cm{S}, v \in \cm{S}}\phi(u) P_{u, v} = \sum_{u\in \bar{\cm{S}}, v\in \cm{S}} \phi(u) P_{u, v} = |\partial_{in}(\cm{S})|\\
\end{split}
\end{equation}

Therefore, $|\partial_{out}(\cm{S})| = vol(\cm{S}) - |out(\cm{S}) \cap in(\cm{S})| = |\partial_{in}(\cm{S})|$.

\end{proof}

\subsection{Proof of Theorem \ref{thm: psi}}
\label{pf: psi}
\begin{proof}
    From the proof of Lemma \ref{lem: partial_in = partial_out = partialS}, we have $|\partial_{out}(\cm{S})| \leq out(\cm{S}) = vol(\cm{S}), |\partial_{in}(\cm{S})| \leq in(\cm{S}) = vol(\cm{S})$.
    By definition \ref{df: psi}, $\forall v\in \cm{C}, \frac{\psi_{\cm{C}}(v)}{\phi(v)} = \frac{1}{vol(\cm{V})}$; $\forall v\notin \cm{C}, \frac{\psi_{\cm{C}}(v)}{\phi(v)} = 0$. Therefore,
\begin{equation}
    \psi_{\cm{C}}[k] = \begin{cases}
        \frac{k}{vol(\cm{C})} &, \,\, k\in [0, vol(\cm{C})]\\
        \,\, 1 &, \,\, k\in [vol(\cm{C}), 1]\\
    \end{cases}
\end{equation}
By Lemma \ref{lem: PageRank versus original}, 
\begin{equation}
\begin{split}
    pr(\alpha, \psi_{\cm{C}})[|\partial_{out}(\cm{S})|] &\leq \psi_{\cm{C}}[|\partial_{out}(\cm{S})|] = \frac{|\partial_{out}(\cm{S})|}{vol(\cm{C})}\\
    pr(\alpha, \psi_{\cm{C}})[|\partial_{in}(\cm{S})|] &\leq \psi_{\cm{C}}[|\partial_{in}(\cm{S})|] = \frac{|\partial_{in}(\cm{S})|}{vol(\cm{C})}\\
\end{split}
\label{eq: equation to use 2 in theorem psi proof}
\end{equation}

For one-step random walk and any distribution $s$, $sP(\bar{\cm{C}}) - s(\bar{\cm{C}})$ is the amount of probability pushed to $\bar{\cm{C}}$ subtracted by the amount of probability pushed from $\bar{\cm{C}}$. Therefore,
\begin{equation}
\begin{split}
    sP(\bar{\cm{C}}) - s(\bar{\cm{C}}) &= \sum_{u\in \cm{C}, v\in \bar{\cm{C}}} s(u)P_{u, v} - \sum_{u\in \bar{\cm{C}}, v\in \cm{C}} s(u)P_{u, v} \leq \sum_{u\in \cm{C}, v\in \bar{\cm{C}}} s(u)P_{u, v} \\ 
    &\overset{Definition \ref{df: p(A)}}{=} s(\partial_{out}(\cm{C})) \overset{Lemma \, \ref{lem: pa and pa}}{\leq} s[|\partial_{out}(\cm{C})|]
\end{split}
\label{eq: equation to use 1 in theorem psi proof}
\end{equation}
and
\begin{equation}
\begin{split}
    pr(\alpha, \psi_{\cm{C}})(\bar{\cm{C}}) &= (\alpha \psi_{\cm{C}} + (1-\alpha)pr(\alpha, \psi_{\cm{C}})M)(\bar{\cm{C}}) \\
    & \overset{\psi_{\cm{C}}(\bar{\cm{C}}) = 0}{=} ((1-\alpha)pr(\alpha, \psi_{\cm{C}})M)(\bar{\cm{C}}) \\
    & \overset{M = \frac{1}{2}(I + P)}{=} \frac{(1-\alpha)}{2}pr(\alpha, \psi_{\cm{C}})(\bar{\cm{C}}) + \frac{(1-\alpha)}{2}(pr(\alpha, \psi_{\cm{C}})P)(\bar{\cm{C}}) \\
    & \overset{Equation \, \ref{eq: equation to use 1 in theorem psi proof}}{\leq} \frac{(1-\alpha)}{2}pr(\alpha, \psi_{\cm{C}})(\bar{\cm{C}}) + \frac{(1-\alpha)}{2}(pr(\alpha, \psi_{\cm{C}})(\bar{\cm{C}}) + pr(\alpha, \psi_{\cm{C}})[|\partial_{out}(\cm{C})|]) \\
    & = (1-\alpha)pr(\alpha, \psi_{\cm{C}})(\bar{\cm{C}}) + \frac{(1-\alpha)}{2}pr(\alpha, \psi_{\cm{C}})[|\partial_{out}(\cm{C})|] \\
\end{split}
\end{equation}
Hence,
\begin{equation}
\begin{split}
    \alpha pr(\alpha, \psi_{\cm{C}})(\bar{\cm{C}}) &\leq \frac{(1-\alpha)}{2}pr(\alpha, \psi_{\cm{C}})[|\partial_{out}(\cm{C})|] \\
    & \overset{Equation \, \ref{eq: equation to use 2 in theorem psi proof}}{\leq} \frac{(1-\alpha)}{2}\frac{[|\partial_{out}(\cm{C})|]}{vol(\cm{C})} \\
    & \overset{Lemma \, \ref{lem: partial_in = partial_out = partialS}}{\leq} \frac{(1-\alpha)}{2}\frac{[|\partial \cm{C}|]}{vol(\cm{C})} \\
    & \leq \frac{(1-\alpha)}{2}\frac{[|\partial \cm{C}|]}{\min(vol(\cm{C}), 1-vol(\cm{C}))} \\
    &= \frac{1-\alpha}{2} \Phi(\cm{C})
\end{split}
\end{equation}

\begin{equation}
\therefore pr(\alpha, \psi_{\cm{C}})(\bar{\cm{C}}) \leq \frac{1-\alpha}{2\alpha} \Phi(\cm{C}) \leq \frac{\Phi(\cm{C})}{2\alpha}
\end{equation}

Once we have $pr(\alpha, \psi_{\cm{C}})(\bar{\cm{C}}) \leq \frac{\Phi(\cm{C})}{2\alpha}$, we can obtain $pr(\alpha, \psi_{\cm{C}})(\cm{C}) \geq 1 - \frac{\Phi(\bar{\cm{C}})}{2\alpha}$ and finish the proof of the quadratic optimality of the smallest sweep cut conductance.

\end{proof}

\subsection{Proof of Theorem \ref{thm: PhiD no greater than PhiC}}
\label{pf: PhiD no greater than PhiC}

\begin{proof}
    By definition \ref{df: psi},

\begin{equation}
    \psi_\cm{C}(v) = \begin{cases}
        \frac{\phi(v)}{vol(\cm{C})} & ,\,\, v \in \cm{C}\\
        \,\, 0                           & ,\,\, otherwise\\
    \end{cases}
\end{equation}

\begin{equation}
    \psi_\cm{D}(v) = \begin{cases}
        \frac{\phi(v)}{vol(\cm{D})} & ,\,\, v \in \cm{D}\\
        \,\, 0                           & ,\,\, otherwise\\
    \end{cases}
\end{equation}

\begin{equation}
    \psi_{\cm{C}\setminus\cm{D}}(v) = \begin{cases}
        \frac{\phi(v)}{vol(\cm{\cm{C}\setminus\cm{D}})} & ,\,\, v \in \cm{\cm{C}\setminus\cm{D}}\\
        \,\, 0                           & ,\,\, otherwise\\
    \end{cases}
\end{equation}

Therefore,

\begin{equation}
    \psi_{\cm{C}} = \frac{vol(\cm{D})}{vol(\cm{C})} \psi_\cm{D} + \frac{vol(\cm{C}\setminus\cm{D})}{vol(\cm{C})} \psi_{\cm{C}\setminus\cm{D}}
\label{eq: linear probability decomposition}
\end{equation}

According to the linear property of (lazy) PageRank vectors where $pr(\alpha, s) = \alpha s \sum_{k=0}^\infty ((1-\alpha) M)^k$, described by definition \ref{df: pagerank}, from equation \ref{eq: linear probability decomposition}, we further have
\begin{equation}
\begin{split}
    & \qquad \,\,  pr(\alpha, \psi_{\cm{C}}) = \frac{vol(\cm{D})}{vol(\cm{C})} pr(\alpha, \psi_\cm{D}) + \frac{vol(\cm{C}\setminus\cm{D})}{vol(\cm{C})} pr(\alpha, \psi_{\cm{C}\setminus\cm{D}})\\
    &\Rightarrow pr(\alpha, \psi_{\cm{C}})(\bar{\cm{C}}) = \frac{vol(\cm{D})}{vol(\cm{C})} pr(\alpha, \psi_\cm{D})(\bar{\cm{C}}) + \frac{vol(\cm{C}\setminus\cm{D})}{vol(\cm{C})} pr(\alpha, \psi_{\cm{C}\setminus\cm{D}})(\bar{\cm{C}})
\end{split}
\end{equation}

Since $\frac{vol(\cm{D})}{vol(\cm{C})} + \frac{vol(\cm{C}\setminus\cm{D})}{vol(\cm{C})} = 1$, at least one of the following equations holds.
\begin{equation}
    pr(\alpha, \psi_\cm{D})(\bar{\cm{C}}) \leq pr(\alpha, \psi_{\cm{C}})(\bar{\cm{C}})
\end{equation}

\begin{equation}
    pr(\alpha, \psi_{\cm{C}\setminus\cm{D}})(\bar{\cm{C}}) \leq pr(\alpha, \psi_{\cm{C}})(\bar{\cm{C}})
\end{equation}

Therefore, for any vertex set $\cm{C} \subseteq \cm{V}$, and any of its a subset $\cm{D} \subseteq \cm{C}$. At least one of $pr(\alpha, \psi_\cm{D})(\bar{\cm{C}})$ and $pr(\alpha, \psi_{\cm{C}\setminus\cm{D}})(\bar{\cm{C}})$ is no greater than $pr(\alpha, \psi_{\cm{C}})(\bar{\cm{C}})$. Hence, if $\cm{D}$ is randomly sampled from the subsets of $\cm{C}$, with probability at least $\frac{1}{2}$, $pr(\alpha, \psi_\cm{D})(\bar{\cm{C}}) \leq pr(\alpha, \psi_{\cm{C}})(\bar{\cm{C}})$.

\end{proof}

\subsection{Proof of Theorem \ref{thm: conductance bound}}
\label{pf: conductance bound}
\begin{proof}
    In Theorem \ref{thm: psi}, let $\cm{C} = \cm{S}^{*}$. Then, since $\cm{S} \subseteq \cm{S}^{*}$, from Theorem \ref{thm: PhiD no greater than PhiC}, At least one of $pr(\alpha, \psi_\cm{S})(\bar{\cm{S}^{*}})$ and $pr(\alpha, \psi_{\cm{S}^*\setminus\cm{S}})(\bar{\cm{S}^{*}})$ is no greater than $pr(\alpha, \psi_{\cm{S}^*})(\bar{\cm{S}^{*}})$. As $\cm{S}$ is uniformly randomly sampled from $\cm{V}$, the probabilities of sampling $\cm{S}$ or $\cm{S}^*\setminus\cm{S}$ are the same. Since both $\cm{S}$ and $\cm{S}^*\setminus\cm{S}$ has optimal local cluster $\cm{S}^*$, with probability at least $\frac{1}{2}$, 
\begin{equation}
    pr(\alpha, \psi_{\cm{S}})(\bar{\cm{S}^{*}}) \leq pr(\alpha, \psi_{\cm{S}^*})(\bar{\cm{S}^{*}}) \leq \frac{\Phi(\cm{S}^{*})}{2\alpha} \Rightarrow pr(\alpha, \psi_{\cm{S}})(\cm{S}^{*}) \geq 1 - \frac{\Phi(\cm{S}^{*})}{2\alpha}
\end{equation}

\begin{equation}
    \therefore pr(\alpha, \psi_{\cm{S}})(\cm{S}^{*}) - vol(\cm{S}^{*}) \geq 1 - \frac{\Phi(\cm{S}^{*})}{2\alpha} - vol(\cm{S}^{*})
\end{equation}

According to Theorem \ref{thm: conductance_satisfying_shrunk}, $\forall z\in [0, \frac{1}{e}]$,  if $\delta = 1 - \frac{\Phi(\cm{S}^{*})}{2\alpha} - vol(\cm{S}^{*}) > z$, then,

\begin{equation}
    \Phi(pr(\alpha, \psi_{\cm{S}})) < \sqrt{\frac{9\alpha \ln\frac{1}{z}}{\delta - z}} = \sqrt{\frac{9\alpha \ln\frac{1}{z}}{1 - \frac{\Phi(\cm{S}^{*})}{2\alpha} - vol(\cm{S}^{*}) - z}}
\end{equation}

Let $\alpha = \Phi(\cm{S}^{*}) \in [0, 1]$, then,

\begin{equation}
    \Phi(pr(\alpha, \psi_{\cm{S}})) < \sqrt{\frac{9\alpha \ln\frac{1}{z}}{\delta - z}} = \sqrt{\frac{9\Phi(\cm{S}^{*}) \ln \frac{1}{z}}{\frac{1}{2} - vol(\cm{S}^{*}) - z}}
\end{equation}

holds for any $z\in [0, \frac{1}{e}]$. When $vol(\cm{S}^*) \leq \frac{1}{2}$, we can always find $z \leq \frac{1}{2} - vol(\cm{S}^*)$ to bound $\Phi(pr(\alpha, s))$ such that
\begin{equation}
    \Phi(pr(\alpha, \psi_{\cm{S}})) < O(\sqrt{\Phi(\cm{S}^{*})})
\end{equation}

Especially, when $vol(\cm{S}^*) \leq \frac{1}{3}$, let $z = \frac{1}{18}$, 

\begin{equation}
    \Phi(pr(\alpha, \psi_{\cm{S}})) < \sqrt{\frac{9\Phi(\cm{S}^{*}) \ln 18}{\frac{1}{2} - vol(\cm{S}^{*}) - \frac{1}{18}}} \leq \sqrt{81\Phi(\cm{S}^{*}) \ln 18} \leq \sqrt{235\Phi(\cm{S}^{*})}
\end{equation}
\end{proof}

\section{Algorithm Complexity}
\label{ap: algorithm analysis}

\subsection{Worst-case Time Complexity of HyperACL}
\label{sec: time complexity hyper l}
The pseudo-code of HyperACL is given in Algorithm \ref{alg: local graph}. Assume we have direct access to the support of each $\gamma_e$.
Assume the number of hyperedge-vertex connections is $m$, which is the sum of the sizes of the support sets of all $\gamma_e$. Additionally, we assume that the returned local cluster has size $k$.

The computation of $R$ and $W$ takes $O(m)$. The constructed $R$ and $W$ both have $m$ non-zero entries. The construction of $D_{\cm{V}}$ takes $O(|\cm{V}|)$ and the construction of $D_{\cm{E}}$ takes $O(|\cm{E}|)$. Given that each hyperedge has at least 2 vertices and each vertex has at least one hyperedge incident to it, step 1 takes $O(m)$.

Given that $W$ and $R$ both have $m$ non-zero elements, the multiplication $P = D_\cm{V}^{-1}WD_\cm{E}^{-1}R$ takes $O(m^2)$ using CSR format sparse matrix. Therefore step 2 takes $O(m^2)$.

Computing the stationary distribution $\phi$ of $P$ takes $O(|\cm{V}|^2)$ using power iteration. Therefore, step 1, step 2, and step 3 take $O(m^2)$.

Step 4 takes $O(|\cm{V}|^2)$ to calculate the PageRank vector by power iteration. Step 5 takes $O(|\cm{V}| log (|\cm{V}|))$.

In Step 6, for a local cluster of size $l$, computing its conductance takes $O(l(|\cm{V}|-l)$. We record the smallest conductance along computing. Therefore, Steps 6 and 7 take 

\begin{equation}
    \sum_{l=1}^k l\cdot (|\cm{V}| - l) = \sum_{l=1}^k l |\cm{V}| - \sum_{l=1}^k l^2 = \frac{k(k+1)}{2}|\cm{V}| - \frac{k(k+1)(2k+1)}{6} \in O(k^2 |\cm{V}|)
\end{equation}

Note that $m \geq |\cm{V}|$. Therefore, the worst-case time complexity of HyperACL is

\begin{equation}
    O(m^2) + O(|\cm{V}|^2) + O(|\cm{V}| log (|\cm{V}|)) + O(k^2 |\cm{V}|) \in O(m^2 + |\cm{V}|^2 + k^2 |\cm{V}|)
\end{equation}

For GeneralACL, the computation of the transition matrix may not be $P = D_\cm{V}^{-1}WD_\cm{E}^{-1}R$ but depends on specific data and task properties. Assuming the complexity of computing $P$ is $C$, then the worst-case time complexity of GeneralACL is

\begin{equation}
    O(C) + O(|\cm{V}|^2) + O(|\cm{V}|^2) + O(|\cm{V}| log (|\cm{V}|)) + O(k^2 |\cm{V}|) \in O(C + |\cm{V}|^2 + k^2 |\cm{V}|)
\end{equation}

\subsection{Worst-case Space Complexity of HyperACL}
\label{sec: space complexity hyper l}
We make the same assumption as in section \ref{sec: time complexity hyper l}.
The storage of $R, W, D_{\cm{V}}, D_{\cm{E}}$ takes $O(m)$. The storage of $P$ takes $O(|\cm{V}|^2)$. The storage of stationary distribution and the intermediate results takes $O(|\cm{V}|)$. Other steps are not as space-consuming as these.

Therefore, the worst-case space complexity for HyperACL is 
\begin{equation}
    O(m)+ O(|\cm{V}|^2) + O(|\cm{V}|) \in O(m + |\cm{V}|^2)
\end{equation}

For GeneralACL, Assuming the space complexity of computing $P$ is $C$. The storage of $P$ takes $O(|V|^2)$ and the worst-case space complexity for GeneralACL is

\begin{equation}
    O(C)+ O(|\cm{V}|^2) + O(|\cm{V}|) \in O(C + |\cm{V}|^2)
\end{equation}

\section{Additional Experimental Details}
\label{ap: exp_details}
\input{z4_exp_details}

\end{document}

%% file: logical_flow.tex
\begin{figure*}[t]
\centering
\caption{Logical flow of this work. "A$\rightarrow$B" means A is required for developing B. "Def" stands for definition to save space.}
\label{fig: technical map}
\vspace{3mm}
\resizebox{0.99\linewidth}{!}{%
\begin{forest}
for tree={
  draw, rounded corners, align=center, inner sep=3pt,
  s sep=8mm, l sep=10mm, 
  draw=orange,
  edge={-Latex}
}
[Assumption \ref{assumption: well-defined transition matrix}; \ref{assumption: hypergraph connectivity} (connectivity and transition matrix) \\
Definition \ref{df: graph}; \ref{df: graph random walk} (Graph; Graph Random Walk); Definition \ref{df: p(S)} (Probability of a set)  \\
Definition \ref{df: edvw hypergraph}; \ref{df: R, W, Dv, De}; \ref{df: transition matrix}; \ref{df: stationary distribution} (EDVW Hypergraph; Hypergraph Random Walk; Stationary Distribution) \cite{DBLP:conf/icml/ChitraR19} \\
Theorem \ref{thm: equality of in and out}; Lemma \ref{lem: partial_in = partial_out = partialS} (Equivalence of walking into or walking out of a vertex set) 
  [Def \ref{df: volume of boundary}; \ref{df: volume of a set} (Hypergraph Volumes) \\ Def \ref{df: general graph conductance}; \ref{df: conductance}; Theorem \ref{thm: Phi(S) bound} (Conductance), name=volumesconductance
  ]
  [Definition \ref{df: pagerank}; Lemma \ref{lem: equivalent of lazy and standard} \\ (Lazy Personalized PageRank) \\ Def \ref{df: indicator function and sspr} (Lazy single-source PPR), name=lazyppr]
  [Definition \ref{df: sweep}; \ref{df: optimal conductance of a distribution} (Sweep sets \\ and Optimal conductance), name=sweepset
    [Definition \ref{df: LSC} (Lovász-Simonovits Curve) \\ Definition \ref{df: p(u, v)} ($p(u;v)$ notation) \\ Definition \ref{df: p(A)} ($p(\mathcal{A})$ notation) \\ Definition \ref{df: |A|} ($|\mathcal{A}|$ notation) \\ Definition \ref{df: in out} ($in(\cm{S})$ and $out(\cm{S})$ notations), name=lsc
        [Lemma \ref{lem: ps and pvols}; Lemma \ref{lem: pa and pa} \\ (Properties of Lovász-Simonovits Curve), name=propertylsc
            [Lemma \ref{lem: pvol pvol+-partial} (Inequality of breakpoints \\ on Lovász-Simonovits Curve), name=breakpointlsc, calign=first
                [Theorem \ref{thm: either or} (Upper Bound of the \\ Disagreement: ``either or'' Theorem), name=eitheror]
                [Theorem \ref{thm: conductance_satisfying_shrunk} (Maximum Conductance \\ Given Lovász-Simonovits Lower Bound), name=maximumconductance, no edge]
          ]
        ]
        [Lemma \ref{lem: pms in out}; Lemma \ref{lem: ababab}; Lemma \ref{lem: in out vol}; Lemma \ref{lem: in out partial}; Lemma \ref{lem: in out vol and partial}\\ (Equalities), name=equalities]
    ]
  ]
  [Definition \ref{df: partial in out} \\ ($\partial_{in}(\cm{S})$ and $\partial_{out}(\cm{S})$ notation), name=partialin]
  [Definition \ref{df: psi} \\ (PageRank Starting Distribution), name=pagerankstart
    [Theorem \ref{thm: psi} (Upper-bound PPR with conductance), name=pprupperbound]
    [Theorem \ref{thm: PhiD no greater than PhiC} (Sampling certainty \\ with at least $\frac{1}{2}$ probability), name=sampling
        [Theorem \ref{thm: conductance bound} (\textbf{Quadratic Optimality}), name=quadraticoptimality]
    ]
  ]
]
\draw[-Latex,black] (equalities) .. controls +(south:14mm) and +(east:30mm) .. (breakpointlsc);
\draw[-Latex,black] (volumesconductance) .. controls +(south:14mm) and +(west:40mm) .. (eitheror);
\draw[-Latex,black] (breakpointlsc) .. controls +(north east:14mm) and +(south west:70mm) .. (pprupperbound);
\draw[-Latex,black] (pprupperbound) .. controls +(south east:14mm) and +(north west:20mm) .. (quadraticoptimality);
\draw[-Latex,black] (maximumconductance) .. controls +(east:40mm) and +(south west:70mm) .. (quadraticoptimality);
\draw[-Latex,black] (eitheror.east) -- (maximumconductance.west);
\end{forest}
}
\end{figure*}

%% file: z4_exp_details.tex
\newcommand{\zhuang}[1]{{\textsf{\textcolor{orange}{[To be filled by Hengyu]}}}}

\subsection{Environments}
We run all our experiments on a Windows 11 machine with a 13th Gen Intel(R) Core(TM) i9-13900H CPU, 64GB RAM, and an NVIDIA RTX A4500 GPU. One can also run the code on a Linux machine. All the code of our algorithms is written in Python. The Python version in our environment is 3.11.4. In order to run our code, one has to install some other common libraries, including PyTorch, pandas, numpy, scipy, and ucimlrepo. Please refer to our README in the code directory for downloading instructions.

\subsection{Datasets}
\label{ap: datasets}
Table \ref{tb: datasets local} summarize the statistics and attributes of each dataset we used in the experiments.

For local clustering, We first download the newest AMiner DBLP v14 from its official website\footnote{\url{https://www.aminer.org/citation}}, then filter out the publications with incomplete information, then retrieve all the papers from the year 2018 (inclusive) to the year 2023 (exclusive). Based on this, we retrieve all publications in venues ICML, NIPS, and ICLR to construct DBLP-ML; all publications in venues CVPR, ECCV, and ICCV to construct DBLP-CV; all publications in venues ACL, EMNLP, and NAACL to construct DBLP-NLP; all publications in venues KDD, SIGIR, WWW to construct DBLP-IR. Our datasets are provided within the code of this work.

When filtering out the publications with incomplete information, we remove the publications that have one of the following: (1) any author with an unknown organization. (2) any author with an unknown author ID. The author ID, which is a unique string for each author, is different from the vertex ID. Given the fact that lots of publications have missing information, the final datasets that we are using contain only a portion of actual publications in the venues.

According to the statistics from Paper Copilot, from the year 2018 to the year 2023, 4895 papers were accepted by ICML\footnote{\url{https://papercopilot.com/statistics/icml-statistics/}}, 3481 papers were accepted by ICLR\footnote{\url{https://papercopilot.com/statistics/iclr-statistics/}}, 9344 papers were accepted by NeurIPS\footnote{\url{https://papercopilot.com/statistics/neurips-statistics/}}. The total number of vertices in the DBLP-ML datasets should be 17720, and we have 6617 ($\approx 37.3\%$) in our dataset.

According to the public GitHub repository about conference acceptance\footnote{\url{https://github.com/lixin4ever/Conference-Acceptance-Rate}}, from the year 2018 to the year 2023, 7471 papers were accepted by CVPR, 2694 papers were accepted by ICCV, and 3782 papers were accepted by ECCV. The total number of vertices in the DBLP-CV dataset should be 13947, and we have 4287 ($\approx 31\%$). 

According to the same GitHub repository about conference acceptance, from the year 2018 to the year 2023, 3151 papers were accepted by ACL, 3868 papers were accepted by EMNLP, 1395 papers were accepted by NAACL. The total number of vertices in the DBLP-NLP dataset should be 5991, and we have 1728 ($\approx 29\%$) in our data set.

According to the same GitHub repository about conference acceptance, from the year 2018 to the year 2023, 1215 papers were accepted by KDD, 629 papers were accepted by SIGIR, and 1293 papers were accepted by WWW. The total number of vertices in the DBLP-IR dataset should be 3137, and we have 2848 ($\approx 91\%$). There is no author in UCB that coauthors any paper in our DBLP-IR (actually, according to CS Rankings\footnote{\url{https://csrankings.org/\#/fromyear/2018/toyear/2022/index?inforet\&us}}, only 4 authors and 5 papers were accepted into KDD, SIGIR or WWW from the year 2018 to the year 2023). Therefore for our DBLP-IR, instead of first uniformly sampling one seed institute from \{MIT, CMU, Stanford, UCB\}, we uniformly sample one from \{MIT, CMU, Stanford\}.

Table \ref{tb: datasets local} shows the statistics of our datasets used in the local clustering experiments. The meanings of columns are the name of the hypergraphs, number of vertices/authors, number of hyperedges/publications, number of authorship connections, number of edges in the clique expansion graphs, number of organizations, number of authors affiliated with Massachusetts Institute of Technology, number of authors affiliated with Carnegie Mellon University, number of authors affiliated with Stanford University, number of authors affiliated with UC Berkeley, and the included venues.

\textbf{Edge-Dependent Vertex Weight Assignment on Citation Hypergraphs for Local Clustering.}
For a publication in computer science, usually, the middle authors have fewer contributions to the paper than the authors in the front (who are usually the principal investigators) and the authors in the back (who are usually advisors who manage the project). Therefore, for each publication/hyperedge, we exponentially assign more vertex weights for the authors/vertex in the front or in the back. Specifically, 

(1) For a publication $e$ that has an odd number ($2k+1$) of authors indexed by $0, 1, ..., 2k$, we assign vertex weight $1$ for the middle author (indexed by $k$), and multiply the vertex weight by $2$ when sweeping to the front or the back:

\begin{equation}
    \gamma_e(x) = 2^{|x-k|}, \forall x \in \{0, 1, ..., 2k\}
\end{equation}

(2) For a publication $e$ that has an even number ($2k$) of authors indexed by $0, 1, ..., 2k-1$, we assign vertex weight $1$ for the middle two authors (indexed by $k-1$, $k$), and multiply the vertex weight by $2$ when sweeping to the front or the back:

\begin{equation}
    \gamma_e(x) = \begin{cases}
        2^{k-1-x}, &\forall x \in \{0, 1, ..., k-1\}\\
        2^{x-k},   &\forall x \in \{k, k+1, ..., 2k\}
    \end{cases}
\end{equation}

\begin{table}[h]
\centering
\caption{Statistics of Constructed Hypergraphs in Local Clustering Experiments}
\scalebox{0.7}{
\renewcommand\arraystretch{2}
\begin{tabular}{l l l l l l c c c c l}
\hline
Hypergraph  & $|\cm{V}|$   & $|\cm{E}|$     & $\sum_{v\in\cm{V}} E(v)$   & $|\cm{E}|$ clique & \# orgs & MIT & CMU & Stanford & UCB  & included venues \\ \hline
DBLP-ML     & 14958         & 6617          & 25790         & 83462 & 5117 & 462 & 409 & 478 & 520 & ICML, NeurIPS, ICLR\\ 
DBLP-CV     & 12116         & 4287          & 19883         & 76360 & 3813 & 115 & 175 & 128 & 135 & CVPR, ECCV, ICCV\\
DBLP-NLP    & 4555          & 1728          & 6907          & 24346 & 1165 & 42  & 146 & 45  & 32  & ACL, EMNLP, NAACL\\
DBLP-IR     & 8601          & 2848          & 12203         & 48965 & 3409 & 18  & 72  & 66  & 0 & KDD, SIGIR, WWW\\
\hline
\end{tabular}
\label{tb: datasets local}
}
\end{table}

\renewcommand\arraystretch{2}

\subsection{Metrics}
\label{ap: metrics}

The F1 score between two sets $\mathcal{A}_m$ and $\mathcal{A}_c$ is defined as

\begin{equation}
\begin{split}
    &\textit{TP(True Positive)} = |\mathcal{A}_{m} \cap \mathcal{A}_{c}|\\
    &\textit{FP(False Positive)} = |\mathcal{A}_{m} \setminus \mathcal{A}_{c}|\\
    &\textit{FN(False Negative)} = |\mathcal{A}_{c} \setminus \mathcal{A}_{m}|\\
    &\textit{precision} = \frac{TP}{TP + FP}\\
    &\textit{recall} = \frac{TP}{TP + FN}\\
    &\textit{F1} = \frac{2 \times \textit{precision} \times \textit{recall}}{\textit{precision} + \textit{recall}}
\end{split}
\label{eq: f1 score}
\end{equation}

Assume for an observation, all the authors in the academic institute we sampled construct vertex set $\cm{A}_c$, and the algorithm returns a local cluster $\cm{A}_m$. We compute the conductance of vertex set $A_m$ as Definition \ref{df: conductance}. 
\begin{equation}
    \Phi(\cm{A}_m) = \frac{|\partial\cm{A}_m|}{\min(vol(\cm{A}_m), vol(\bar{\cm{A}_m}))}\\ = \frac{|\partial\cm{A}_m|}{\min(vol(\cm{A}_m), 1 - vol(\cm{A}_m))}
\end{equation}
Then, we compute the F1 score between $A_m$ and $A_c$ as Equation \ref{eq: f1 score}.

\subsection{Baselines}
\label{ap: baselines}

\textbf{CLIQUE++}. For each hyperedge $e$, each $u, v \in e$ with $u \neq v$, we add an edge $uv$ of weight $w(e)$. Then, we apply the ACL algorithm \cite{DBLP:conf/focs/AndersenCL06} to compute the sweep cuts of the weighted clique graph. We use exact PageRank vector instead of an approximated one, and adopt the same early-stop mechanism described at the end of section \ref{section: local}.

\textbf{STAR++}. For each hyperedge $e$, we introduce a new vertex $v_e$. For each vertex $u \in e$, we add an edge $uv_e$ of weight $w(e)/|e|$. After converting the hypergraph into a star graph, we do the same algorithm for local clustering as in CLIQUE++.

\textbf{DiffEq} \cite{DBLP:conf/kdd/Takai0IY20}. We directly use the official code\footnote{\url{https://github.com/atsushi-miyauchi/Hypergraph_clustering_based_on_PageRank}} of this algorithm. This method sweeps over the sweep sets obtained by differential equations for local clustering. For global partitioning, it simply calls local clustering for every vertex and returns the best in terms of conductance. Originally, this algorithm could only take one starting vertex for local clustering. We modified the code to add one additional vertex that connects the 5 starting vertices in each observation. Then regard the newly added vertex as the starting vertex, so that it can obtain the local cluster for the given 5 starting vertices.

In local clustering experiments, since in Theorem \ref{thm: conductance bound}, $\alpha = \Phi(\cm{S}^*)$ is agnostic because the obtaining the optimal $\cm{S}^*$ is NP. We run our HyperACL two times. For the first time, we take $\alpha = \Phi(\cm{S})$, the conductance of the starting vertex set, and obtain a local cluster $\cm{S}'$. Then, we take $\alpha = \Phi(\cm{S}')$ and run the algorithm again to obtain a local cluster as the result. For CLIQUE++ and STAR++, we directly pass $\Phi(\cm{S}')$ obtained from HyperACL to them as the $\alpha$ value. For a fair comparison, we do not time the first execution of HyperACL, which obtains a good $\alpha$ for all algorithms to use.

\section{Additional Experiment Results}
\label{ap:additional_exp}

\subsection{HyperACL on Trivago, NTU2012 and ModelNet}
\label{ap:more_datasets}
We run HyperACL on three additional datasets. Trivago \cite{chodrow2021generative} is an e-commerce hotel-search hypergraph, where each node is a hotel and each hyperedge corresponds to a user session. NTU2012 \cite{wu20153d} and ModelNet \cite{chen2003visual} are from computer vision / graphics domains. These additions substantially broaden the evaluation beyond citation networks. We report results in Table \ref{tab:hyperacl_results}. Overall, these results show that HyperACL generalizes beyond citation networks and consistently outperforms the baselines.

These additional results also verify the scalability of HyperACL on large hypergraphs. The Trivago dataset has 172,738 nodes, 10 times larger than the datasets in the main pages. On an NVIDIA A40 GPU, the average runtime is 14.88 seconds for each clustering run. Notably, the baseline methods CLIQUE++ and STAR++ both run out of memory and do not finish on this dataset. These results provide direct evidence that HyperACL remains practical on substantially larger hypergraphs.

\begin{table}[t]
\centering
\caption{Performance comparison on Trivago, NTU2012, and ModelNet datasets.}
\label{tab:hyperacl_results}
\resizebox{\linewidth}{!}{
\begin{tabular}{lcccccc}
\toprule
\textbf{Method} 
& \textbf{Trivago-conductance ($\downarrow$)} 
& \textbf{Trivago-F1 ($\uparrow$)} 
& \textbf{NTU2012-conductance ($\downarrow$)} 
& \textbf{NTU2012-F1 ($\uparrow$)} 
& \textbf{ModelNet-conductance ($\downarrow$)} 
& \textbf{ModelNet-F1 ($\uparrow$)} \\
\midrule
HyperACL & 0.1783 & 0.4032 & 0.0886 & 0.7426 & 0.1634 & 0.4658 \\
CLIQUE++ & OOM & OOM & 0.1531 & 0.6712 & 0.2463 & 0.4193 \\
STAR++ & OOM & OOM & 0.2036 & 0.4886 & 0.2215 & 0.4487 \\
\bottomrule
\end{tabular}
}
\end{table}

\subsection{GeneralACL results}
\label{ap:graph_datasets}

Although this work primarily focuses on hypergraph local clustering, each hypergraph experiment can naturally induce a corresponding graph experiment by reducing hyperedges to pairwise graph connections. To further examine the applicability of GeneralACL in standard graph settings, we additionally evaluate it on the Bitcoin-Alpha graph dataset and compare it with the original ACL baseline.

\begin{table}[h]
\centering
\caption{Graph local clustering results on Bitcoin-Alpha. Lower conductance and higher F1 indicate better performance.}
\label{tab:bitcoin_alpha_graph}
\begin{tabular}{lcc}
\toprule
\textbf{Method} & \textbf{Conductance} ($\downarrow$) & \textbf{F1} ($\uparrow$) \\
\midrule
GeneralACL & 0.151 & 0.063 \\
ACL        & 0.152 & 0.059 \\
\bottomrule
\end{tabular}
\end{table}

As shown in Table~\ref{tab:bitcoin_alpha_graph}, GeneralACL achieves slightly lower conductance and higher F1 than ACL. These results suggest that the proposed generalized formulation remains effective in graph settings and can better leverage the available graph structure. Together with the additional hypergraph experiments reported in Appendix~\ref{ap:more_datasets}, this further supports the robustness of GeneralACL across both graph and hypergraph datasets.

\subsection{Ablation Study of Early Stop}

\subsubsection{Does Early Stop Help?}
\label{ap:early_stop}

We first study whether early stopping improves the practical efficiency of GeneralACL. On Trivago, the algorithm does not finish within one hour without early stopping. With early stopping, the average total runtime is reduced to 14.88 seconds. Among this runtime, 10.82 seconds are spent on the sweep procedure, indicating that sweep remains the main computational bottleneck even after early stopping is applied.

We further evaluate the effect of early stopping on NTU-2012, where the full algorithm without early stopping can still finish. As shown in Table~\ref{tab:early_stop_effect}, early stopping substantially reduces runtime from 178.809 seconds to 25.861 seconds, while the conductance only changes from 0.0725 to 0.0731. This suggests that early stopping provides a significant acceleration with almost no degradation in clustering quality.

\begin{table}[h]
\centering
\caption{Effect of early stopping on NTU-2012. Lower conductance and runtime are better.}
\label{tab:early_stop_effect}
\begin{tabular}{lcc}
\toprule
\textbf{Setting} & \textbf{Conductance}($\downarrow$) & \textbf{Runtime (s)}($\downarrow$) \\
\midrule
Without early stopping & 0.0725 & 178.809 \\
With early stopping    & 0.0731 & 25.861 \\
\bottomrule
\end{tabular}
\end{table}

\subsubsection{Early-Stop Hyperparameter Study}
\label{ap:early_stop_hyper}

We further conduct an ablation study on the early-stop hyperparameter using DBLP-ML. Specifically, we vary the number of early-stop steps and report the resulting conductance and runtime in Table~\ref{tab:early_stop_hyper}. The results show a clear quality-efficiency trade-off. Using fewer early-stop steps leads to faster execution but worse conductance, while increasing the number of steps improves clustering quality at the cost of longer runtime.

\begin{table}[h]
\centering
\caption{Ablation study of the early-stop hyperparameter on DBLP-ML. Lower conductance and runtime are better.}
\label{tab:early_stop_hyper}
\begin{tabular}{ccc}
\toprule
\textbf{Early-Stop Steps} & \textbf{Conductance}($\downarrow$) & \textbf{Runtime (s)}($\downarrow$) \\
\midrule
10 & 0.2046 & 0.9072 \\
20 & 0.1848 & 1.4950 \\
50 & 0.1586 & 3.5840 \\
\bottomrule
\end{tabular}
\end{table}

\subsection{Ablation of Different Weighting Schemes}
\label{ap:weighting_scheme}

We study the sensitivity of GeneralACL to different edge-dependent vertex weighting schemes on DBLP-ML. Specifically, we vary the weighting base and report the resulting conductance and runtime in Table~\ref{tab:weighting_scheme}. The results show that GeneralACL remains stable under different weighting bases. In particular, base 2 achieves the best conductance, while all three settings yield comparable runtime. This suggests that the proposed method is robust to the choice of weighting scheme.

\begin{table}[h]
\centering
\caption{Ablation study of different edge-dependent vertex weighting schemes on DBLP-ML. Lower conductance and runtime are better.}
\label{tab:weighting_scheme}
\begin{tabular}{ccc}
\toprule
\textbf{Weighting Base} & \textbf{Conductance}($\downarrow$) & \textbf{Runtime (s)}($\downarrow$) \\
\midrule
2 & 0.1590 & 0.1125 \\
3 & 0.1773 & 0.1012 \\
4 & 0.1689 & 0.1077 \\
\bottomrule
\end{tabular}
\end{table}

\section{Discussions}
\label{ap:discussions}

\subsection{Technical Challenges Beyond Classic ACL}
\label{ap:technical_challenges}

We discuss the main technical challenges in extending classic ACL to the generalized setting considered in this work.

First, the generalized setting requires redefining the basic quantities used in ACL beyond undirected graphs. In classic ACL, local clustering is naturally characterized through node degrees, edge counts, volume, and conductance. In our setting, these quantities are no longer directly available in their original form due to the presence of hyperedges, edge-dependent vertex weights, and directionality. A key challenge is therefore to develop a unified formulation of volume, cut, and conductance that is consistent across these different graph structures. This is not merely a change of notation, but the foundation that allows ACL-style local clustering to be studied under one common framework.

Second, the analysis needs to move from a discrete to a more general continuous setting. In classic ACL, many arguments rely on integer-valued quantities induced by unweighted graph edges. In our setting, edge-dependent weights and directed interactions make the corresponding quantities real-valued. As a result, the original discrete arguments cannot be applied directly, and the proof requires a substantially revised analysis over continuous quantities.

Third, our theoretical guarantee is more general than the original ACL result in several aspects. The classic ACL theorem requires the target cluster to be globally optimal among all subsets. In contrast, our result only requires the target cluster to be optimal among sets containing the given seed set, or among sets containing the complementary seed set inside the target cluster. Moreover, the original probabilistic guarantee is stated for a single starting vertex, while our result extends the guarantee to multi-seed starting sets.

\subsection{Intuition Behind the Optimality Condition}
\label{ap:optimality_condition_intuition}

We further discuss the optimality condition in Theorem~\ref{thm: conductance bound}. The condition requires that the target cluster $S^*$ containing the seed set $S$ should also remain optimal when the algorithm is seeded from $S^* \setminus S$. Intuitively, this condition is expected to hold when vertices inside $S^*$ are strongly connected to each other, while connections from $S^*$ to the outside are sparse. In this case, starting from $S^* \setminus S$ should still recover the same cluster $S^*$, since the original seed vertices in $S$ remain well tied to $S^* \setminus S$, whereas vertices outside $S^*$ are less likely to be absorbed into the low-conductance cluster.

A simple extreme case is when $S^*$ is disconnected from the rest of the graph. Then any nonempty seed subset inside $S^*$ can recover $S^*$ as an optimal cluster with zero conductance. More generally, the condition captures the stability of the target cluster under different seed choices within the same coherent local community.

The probability factor $1/2$ comes from the symmetry in Theorem~\ref{thm: PhiD no greater than PhiC}. For the two seed sets $S$ and $S^* \setminus S$, at least one satisfies the desired inequality. Therefore, without prior knowledge of which side is better, uniformly choosing between them yields a natural worst-case probability guarantee of $1/2$. In practice, local clustering seeds are usually selected meaningfully rather than uniformly at random, often from representative or dense regions of the target cluster. In such cases, the empirical success probability can be higher than this worst-case bound.

\subsection{Discussion on Approximate Personalized PageRank}
\label{ap:approx_ppr_discussion}

Our implementation computes personalized PageRank directly rather than using approximate personalized PageRank. This design choice is mainly motivated by two reasons. First, many practical approximation strategies already exist for PageRank-style computation, and our focus is on the generalized local clustering formulation rather than engineering a specific approximation routine. Second, our method still needs the global stationary distribution, and our experiments show that the PageRank computation itself is not the dominant computational bottleneck compared with the sweep procedure.

Nevertheless, approximate personalized PageRank can also be extended to our generalized setting. Since the proposed formulation preserves the random-walk interpretation under the generalized transition structure, standard push-style approximation ideas can be adapted by maintaining residual mass and only propagating residuals above a prescribed threshold. This would further reduce computation on large-scale graphs and hypergraphs, while preserving the same local-computation spirit as classic ACL. We leave a more systematic study of approximate solvers for future work.